\documentclass[10pt]{article}

\usepackage{graphicx}
\usepackage[round]{natbib}
\usepackage{hyperref}
\hypersetup{
    colorlinks = true,
    urlcolor   = blue,
    citecolor  = black,
}

\usepackage{amsmath}
\usepackage{authblk}
\usepackage{booktabs}
\usepackage[noabbrev]{cleveref}
\usepackage[T1]{fontenc}
\usepackage[cm]{fullpage}
\usepackage{xcolor}
\usepackage{siunitx}
\usepackage{url}

\numberwithin{equation}{section}

\title{\bf The influence of freestream turbulence on the development of a wind turbine wake}
\author[1]{Stefano Gambuzza\footnote{Current address: School of Engineering, Institute for Energy Systems, University of Edinburgh, Edinburgh EH9 3FB, United Kingdom. Email: \href{mailto:s.gambuzza@ed.ac.uk}{s.gambuzza@ed.ac.uk}}}

\author[1]{Bharathram Ganapathisubramani}
\affil[1]{Aerodynamics and Flight Mechanics Research Group, \par \noindent University of Southampton, Southampton SO17 1BJ, United Kingdom}

\date{}

\newcommand{\uinf}{\ensuremath{U_\infty}}
\newcommand{\red}{\ensuremath{Re}}
\newcommand{\iinf}{\ensuremath{I_\infty}}
\newcommand{\ifilt}{\ensuremath{I_\mathrm{filt}}}
\newcommand{\tscale}{\ensuremath{T_0}}
\newcommand{\tsr}{\ensuremath{\lambda}}
\newcommand{\ct}{\ensuremath{C_T}}
\newcommand{\cp}{\ensuremath{C_P}}
\newcommand{\kfit}{\ensuremath{k_\mathrm{fit}}}

\begin{document}
\maketitle

\begin{abstract}
The wake of an isolated model-scale wind turbine is analysed in a set of inflow conditions having freestream turbulence intensity between \SI{3}{\percent} and \SI{12}{\percent}, and integral time scales in the range of 0.1 to 10 times the convective timescale based on the turbine diameter. It is observed that the wake generated by the turbine evolves more rapidly, with the onset of the wake evolution being closer to the turbine, for high turbulence intensity and low integral time scale flows, in accordance with literature, while flows at higher integral time scales result in a slow wake evolution, akin to that generated by low-turbulence inflow conditions despite the highly-turbulent ambient condition.
The delayed onset of the wake evolution is connected to the stability of the shear layer enveloping the near wake, which is favoured for low-turbulence or high-integral time scale flows, and to the stability of the helical vortex set surrounding the wake, as this favours interaction events and prevents momentum exchange at the wake boundary which hinder wake evolution.
The rate at which the velocity in the wake recovers to undisturbed conditions is instead analytically shown to be a function of the Reynolds shear stress at the wake centreline, an observation that is confirmed by measurements. The rate of production of Reynolds shear stress in the wake is then connected to the power harvested by the turbine to explain the differences between flows at constant turbulence intensity and different integral time scales.
\end{abstract}


\section{Introduction} \label{sec:intro}
Wind turbines are machines that present a number of peculiar characteristics regarding their operation: firstly, these are situated in the lowermost portion of the atmospheric boundary layer, and thus harvest power from turbulent, sheared inflows.
Moreover, to accommodate for the rising market demand for renewable energy, these are often grouped in wind farms that house a large number of these machines in a finite area: as a result, all turbines that are not situated in the first row of the wind farm experience, as incoming flow, a combination of the wakes of upstream machines and turbulent freestream entrained from the atmospheric boundary layer surrounding the farm.
In particular, the presence of upstream machines limits the power that a wind turbine in the back rows of a farm can harvest, as it will harvest energy from a lower-momentum inflow \citep{Frandsen2002,Barthelmie2010a}.
For this reason, a good knowledge of the wake generated by a wind turbine in a complex, turbulent inflow is paramount to the prolonged development and deployment of wind energy around the world.

As mentioned, the environment in which wind turbines operate is turbulent and sheared: this is often parametrised with the freestream turbulence intensity, defined as
\begin{equation}
    \iinf{} = \frac{\sqrt{\overline{u'^2}}}{\uinf},
\end{equation}
where \uinf{} is the bulk freestream speed, and $\sqrt{\overline{u'^2}}$ is the standard deviation of the velocity time-history.
\citet{Elliott1990} report data showing that turbines in an on-shore environment are subjec to $\iinf < \SI{15}{\percent}$ for \SI{95}{\percent} of their operating life.
Similarly, \citet{Wagner2011} and \citet{Pena2016} report that turbines at the H\o{}vs\o{}re on-shore testing site experience values of \iinf{} between \SI{2}{\percent} and \SI{14}{\percent}, with a strong dependency on the wind direction: less turbulent inflows are observed for winds arising from the ocean.
Turbulence intensity for off-shore sites is usually lower, with values of \iinf{} comprised between \SI{6}{\percent} and \SI{8}{\percent} \citep{Barthelmie2005,Turk2010}.

Wakes of wind turbines are complex in nature, being characterised by the superposition of a large number of events:
in the simplest of descriptions, the time-averaged wake of a wind turbine is, sufficiently far from the turbine rotor, characterised by a self-similar velocity deficit profile having a Gaussian distribution in the stream-normal directions \citep{Medici2006}; this is in line with the wakes generated by other axisymmetric bluff bodies such as spheres \citep{Uberoi1970} or porous plates \citep{Rind2012,Aubrun2019,Vinnes2022}.
Closer to the turbine, the actual shape of the velocity profile is dominated by the distribution of pressure around the turbine blades: although sometimes the velocity deficit distribution is constant along a radial direction \citep{Medici2006,Mycek2014,Lignarolo2015}, this need not necessarily be, especially for turbines that are not operating at on-design conditions \citep{Vermeer2003,CarbajoFuertes2018,Foti2018a,Dasari2019}.
The transition from these arbitrary profiles to the self-similar Gaussian ones is understood to be a function of the freestream turbulence intensity \citep{Medici2006,Ishihara2018}.
This region of lower velocity is separated from the surrounding freestream by an annular shear layer which is dominated by the presence of a helical vortex structure \citep{Lissaman1979,Crespo1996,Troldborg2010,DeCillis2020}: this is analogous to the classical horseshoe vortex characteristic of finite wings generating lift, and it takes a helical structure due to the rotation of the blades as freestream convects these vortices downstream.
The stability of this structure has been connected to the onset of the wake evolution \citep{Lignarolo2014,Lignarolo2015}, where a strong shear layer enveloping the wake inhibits the onset of wake evolution; moreover, the stability (or lack thereof) is also seen to drive a low-frequency motion of the wake in the stream-normal directions, named wake meandering \citep{Medici2006,Heisel2018,DeCillis2020}, with this motion being favoured by unstable shear layers.
In general, an increase in the freestream turbulence intensity is seen to hasten the transition to a self-similar velocity profile \citep{Medici2006} and a smaller velocity deficit at a given distance from the turbine \citep{Bastankhah2014,Christiansen2005,Chamorro2012}; this is seen from the point of view of a wind farm operator as a larger distance necessary between turbines in an off-shore environment, as the lower freestream turbulence intensity of those sites does not favour wake recovery as much as in on-shore sites \citep{Christiansen2005}.
The stability of the shear layer is also seen to be related to the freestream turbulence intensity, shown by \citet{Sorensen2015} as a relationship between the breakdown of the helical vortex structure and \iinf{}.

It can be understood that not all these phenomena in the wake are of interest to a wind turbine operator: for this reason, engineering applications usually treat the wake statistically in a simplified form, often ignoring some of the aspects here described.
Analytical wake models are engineering tools often used to predict the wake generated by wind turbines; these often relate the flow velocity in the turbine wake to some global parameters, simple to measure or estimate.
The model that has seen the most widespread use in literature in the last years is the Gaussian wake model \citep{Bastankhah2014}: according to this, one has that the velocity deficit in the turbine wake is given by
    \begin{gather}
        \frac{\Delta U}{\uinf} = \frac{\uinf - U}{\uinf} = 
        \left( 1 - \sqrt{1 - \frac{\ct}{8 (\sigma_w/D)^2}} \right)
        \text{exp}\left( -\frac{1}{2(\sigma_w/D)^2} \left(\frac{r}{D}\right)^2\right), \\
        \frac{\sigma_w}{D} = \epsilon + k^* \frac{x}{D},
    \end{gather}
where \uinf{} is the freestream speed, $U$ is the velocity in the wake function of the streamwise distance from the turbine $x$ and the radial distance away from the turbine axis $r$, $D$ is the turbine radius, $C_T$ is the thrust coefficient, $\epsilon$ is a known function of \ct{} and $k^*$ is a parameter related to the spatial growth of the wake.
The effect of turbulence, which has been seen to be for other bluff-bodies that of hastening the wake development and shortening the wake, is often represented as a change in the wake recovery rate $k^*$.
\cite{Niayifar2016} report an elaboration of LES data of the wake generated by a commercial turbine, showing a linear relationship between the freestream turbulence intensity and the wake recovery rate, for a constant value of \ct{}; this is also observed by \citet{CarbajoFuertes2018} with field data on the wake of a turbine acquired with LiDAR.
For the similar Jensen wake model \citep{Jensen1983}, \citet{Pena2016} show that the wake recovery rate is a linear function of the freestream turbulence intensity if one assumes an inflow that is modelled with Monin-Obukhov similarity theory \citep{Monin1954}.
However, some works in recent literature have outlined how the predictions of wake models can be better tuned by introducing one additional parameter: for instance, \citet{Neunaber2022} shows that the accuracy of the predictions of commonly-used wake models can be improved by introducing a virtual origin in their equations.
This practice, commonplace when treating wakes of bluff bodies, is seldom carried out in wind engineering. 
In this work, the authors measure the wakes of two model-scale wind turbines, subject to either a laminar inflow or to the wake of an upstream machine.
Their data shows that the addition of a virtual origin in analytical wake models can improve the quality of the predictions for the wake of downstream machines, hinting to the conclusion that a virtual origin might include information on the inflow conditions.
However, data included in their paper only consists in the wakes generated under either a laminar inflow or the wake of another turbine: given the large range of turbulent flows that turbines experience, it is important to observe whether a virtual origin can improve on the predictions of analytical wake models for an arbitrary inflow condition, and what its physical meaning is.
While a large body of literature has been dedicated to understanding the effects of freestream turbulence intensity on the wake and on analytical prediction, little has been done to characterise the wake developed by a turbine in the presence of flows with different spectral content of turbulence.
It is well established that the distribution of inflow turbulence affects the near-wake and the power a wind turbine generates \citep{Sheinman1992,Tobin2015,Chamorro2015,Deskos2020,Li2020,Gambuzza2021} and the drag generated by turbine simulators \citep{Blackmore2014}, with turbines being more apt at converting velocity fluctuations into power if those are present as lower-frequency contributions.
Most works in literature therefore highlight how a wind turbine acts as a low-pass filter when converting inflow into mechanical power.
It can thus be assumed that the wake generated by a turbine in flows with different representations in the frequency domain are characterised by different scales, and indeed some works report different spectral composition between the freestream and the wake generated by the turbines \citep{Tobin2019,Heisel2018}.
However, whether this leads to different mechanisms of wake development is an assumption that has seldom been tested in literature, and indeed engineering models do assume no effect of the turbulence spectrum on the wake evolution, with parameters such as $k^*$ being only function of \iinf{}.
For this reason, this paper aims to investigate the relationship between the changing inflow conditions that a  turbine is subject to and the mechanisms that regulate and dominate the development of this wake. We follow on our previous work in \citep{Gambuzza2021} and carry out an experimental study to charactertise the wake of a model-scale turbine under different inflow turbulence conditions.

This paper is structured as follows: \cref{sec:method} will briefly present the experimental methodology used to collect the data here presented, along with its shortcomings.
\Cref{sec:results} will present the results obtained: this chapter is further divided in three subsections, which will relate more in detail to the predictions of engineering models (\cref{subsec:results-models}) and to the physics behind the trends observed(\cref{subsec:results-nearwake} and \cref{subsec:results-farwake}).
\Cref{sec:conclusions} will summarise these findings in a concise manner.

\section{Experimental method} \label{sec:method}
This paper reports the measurements of velocity in the wake of a model-scale wind turbine, measured via planar PIV in a wind tunnel equipped with an active turbulence generating grid. This section will expand on the experimental methodology that has been employed to obtain the results reported in the remainder of the paper, outlining the main characteristics and the limitations of the techniques employed.
Some of the techniques utilised in this study are described in more detail in the previous study published in \cite{Gambuzza2021}.

\subsection{Facility} \label{subsec:method-facility}
The experiments have been carried out in the $3 \times 2$ boundary layer wind tunnel at the University of Southampton.
This is a suction wind tunnel having a rectangular cross-section of size \mbox{$0.9 \times 0.6$ \si{\meter\squared}}, and a total usable length of constant cross-section of \SI{4.5}{\meter}.
Flow is driven by a fan placed downstream of the test section, and flow conditioning is carried out by a set of honeycomb meshes upstream of a contraction that leads to the test section.
During the tests here described, the wind tunnel has been equipped with a turbulence-generating active grid, designed to the specifications of \cite{Makita1991}, which is able to generate turbulent flows with different level of shear and freestream turbulence intensities up to \SI{16}{\percent} \citep{Hearst2017,Li2020}; no shear has been generated for the measurements presented in this study.
This grid is composed of 18 stepper motors that independently drive 11 vertical rods and 7 horizontal rods, each moving a set of agitator wings.
The mesh spacing between rods $M$ is \SI{81}{\milli\meter}.
The grid is operated by changing the angular velocity of each rod to a random value in a predetermined interval, with rods allowed to cruise to speed for a limited time before changing direction and speed: this actuation procedure is in detail described by \cite{Poorte2002} as double-random mode.
The active grid is situated at the inlet of the wind tunnel test section, with the rods covering the whole of the wind tunnel cross-section.

During the tests, the wind tunnel fan has been operated to generate a mean freestream speed \uinf{} equal to \SI{8}{\meter\per\second}.
This has been measured by means of a Pitot probe placed $2.5 M$ upstream of the active grid.
To account for the change in \uinf{} along the test section, this has been calibrated to a Pitot placed at the same location as the turbine in an otherwise empty test section.

\subsection{Model-scale turbine} \label{subsec:method-turbine}
The turbine used during these tests is a speed controlled, fixed pitch model-scale wind turbine, having a rotor of diameter $D$ equal to $\SI{0.18}{\meter}$.
For this value of $D$ and the aforementioned constant value of \uinf, the diameter-based Reynolds number of these tests is $\red = 9.6 \times 10^4$.
Data reported by \cite{Chamorro2012} show that this is sufficiently high to attain Reynolds-independent results both in the mean and the second moment of the wake velocity signals.
The turbine rotor has been directly connected to a brushed permanent magnet DC machine that has been used as a generator to brake the rotor while operating.
This has not been connected to other sources of mechanical or electrical power, with the only torque acting on the turbine shaft being the one generated by the freestream on the turbine blades, and the power generated being transformed into heat dissipated by the motor.
The turbine nacelle has been supported by an aluminium mast of diameter \SI{15.75}{\milli\meter}, which has placed the centre of the rotor at the centre of the test section.
The turbine has then been located at a streamwise distance of $36 M = 16 D$ downstream of the active grid.
The geometry of the blade, defined as the distribution of chord and twist along the blade span, is reported in \cref{tab:turbine-geom}, where $r_\mathrm{tip} = D/2$ is the spanwise location of the last section of the blade.
The turbine blade is designed to harvest maximum power from the incoming flow at a tip-speed ratio \tsr{} of 4; this parameter is defined as
\begin{equation}
	\tsr = \frac{\omega \, r_\mathrm{tip}}{\uinf},
	\label{eq:tsr-definition}
\end{equation}
where $\omega$ is the turbine rotor angular velocity.

\begin{table}
	\centering
	\begin{tabular}{lrrr} \toprule
		 $r/r_\mathrm{tip}$ & $c/r_\mathrm{tip}$ & $\beta$ [deg] & Aerofoil \\ \midrule
		 0.13 & 0.13 & 58.20 & NACA 63-418 \\
		 0.20 & 0.27 & 35.58 & NACA 63-418 \\
		 0.30 & 0.26 & 24.13 & NACA 63-418 \\
		 0.40 & 0.24 & 16.76 & NACA 63-418 \\
		 0.50 & 0.22 & 11.77 & NACA 63-418 \\
		 0.60 & 0.21 & 8.22  & NACA 63-418 \\
		 0.70 & 0.19 & 5.58  & NACA 63-418 \\
		 0.80 & 0.17 & 3.55  & NACA 63-418 \\
		 0.90 & 0.16 & 1.94  & NACA 63-418 \\
		 1.00 & 0.14 & 0.64  & NACA 63-418 \\ \bottomrule
	\end{tabular}
	\caption{Turbine geometry, defined as the distribution of chord $c$, twist $\beta$ and aerofoil shape along the blade span coordinate $r$.}
	\label{tab:turbine-geom}
\end{table}

\subsection{Planar particle image velocimetry} \label{subsec:method-piv}
In this study, planar PIV has been employed to measure the velocity in the turbine wake on a stream-parallel vertical plane; the out-of-plane velocity component is not measured by this technique.
The nomenclature that is used in the remainder of the paper defines $x$ to be the streamwise coordinate, positive in the direction of the flow, and $y$ to be the stream-normal, vertical coordinate, positive upwards; both are dimensional and are adimensionalised by the turbine diameter $D$.
The origin of this reference frame is placed at the centre of the turbine rotor, so that $x$ defines the streamwise distance from the rotor-swept plane, and $y$ measures the distance from the turbine axis of rotation.
The convention used here employs the label $u$ for the streamwise component and $v$ for the vertical component of velocity; both are rendered adimensional by the freestream speed \uinf{}.
Moreover, Reynolds decomposition is used to separate the velocity into a time-averaged and a fluctuating, zero-mean component:
\begin{equation}
    u(t) = \overline{u(t)} + u'(t) = U + u'(t),
\end{equation}
where the overline denotes time-averaging, a capital letter represents a quantity constant in time and the prime symbol denotes a signal having zero time-mean.

\begin{figure}[ht!]
    \centering
    \includegraphics{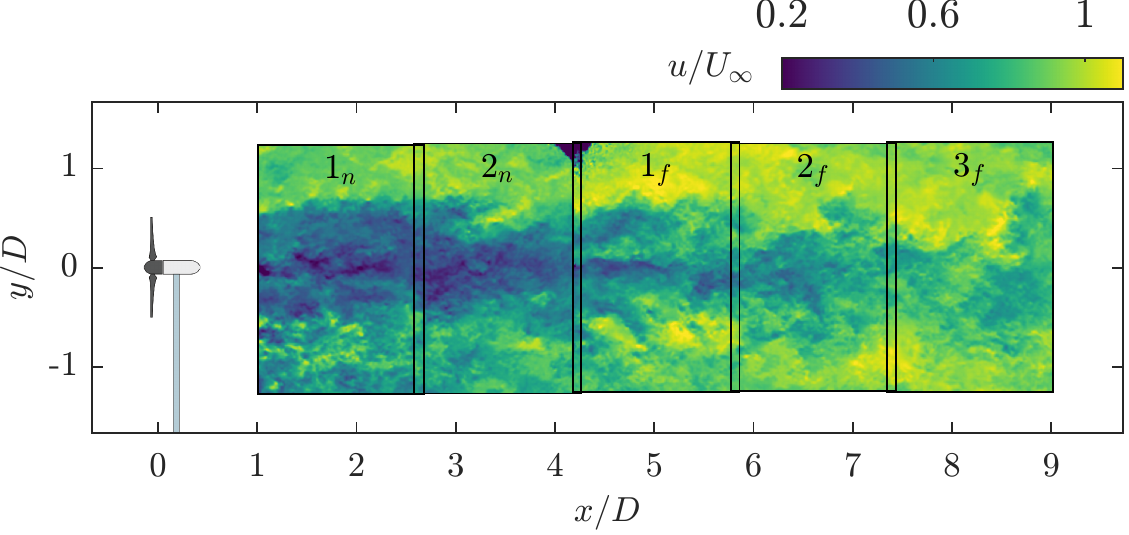}
    \caption{Planar PIV setup with the cameras fields of view in the near wake ($1_n$ and $2_n$) and in the far wake ($1_f$, $2_f$, and $3_f$), including an instantaneous estimate of the streamwise velocity in the turbine wake. All dimensions to scale.}
    \label{fig:piv-setup}
\end{figure}
\Cref{fig:piv-setup} reports a schematic representation of the planar PIV setup used during this study.
Three Imager Pro LX cameras have been used to obtain the five fields of view reported in \cref{fig:piv-setup}, which have been illuminated by a Litron Bernoulli PIV laser: initially, two cameras have simultaneously acquired the velocity field for streamwise distances of 1 to 4.2 rotor diameters downstream of the turbine; these are indicated as positions $1_n$ and $2_n$ in the figure.
Subsequently, these cameras have been moved and a third camera has been added to acquire between 4.1 and 9 diameters of downstream distance; these are locations $1_f$ to $3_f$.
All cameras fields of view have dimensions of \SI{300}{\milli\meter} in the $x$-direction and \SI{450}{\milli\meter} in the $y$-direction in object-plane units, or $1.67 \times 2.5$ rotor diameters; all fields of view overlap the previous and next by a strip \SI{15}{\milli\meter} wide and \SI{450}{\milli\meter} tall, which allows for the statistics fields to be stitched together along the whole span of the measurement domain.
Moreover, instantaneous velocity fields are stitched between cameras $1_n$ and $2_n$, as these have acquired simultaneously, and between cameras $1_f$ to $3_f$ for the same reason.
Processing of the PIV particle displacement snapshots is carried out with LaVision DaVis, specifying an initial window size of $96 \times 96$ px and a final window size of $48 \times 48$ px with a \SI{75}{\percent} overlap between adjacent windows: this results in an overall resolution of 1 velocity vector per \SI{1.1}{\milli\meter} in both the $x$- and $y$-directions, or 162 vectors in one rotor diameter.
Due to the large magnification factor that had to be employed to image the fields of view, tracking particles had a size of approximately 1 px in image-plane units, which lead to the phenomenon of peak-locking as described by \cite{Christensen2004}.
To alleviate the effect of this on the computed statistics, the histogram normalisation correction algorithm presented by \cite{Hearst2015a} has been applied to all correlation maps in image-plane units of displacement, prior to the application of a calibration to convert these into object-plane units of velocity.
The velocity fields have been acquired at a frequency of \SI{0.6}{\hertz}: for a freestream speed of \SI{8}{\meter\per\second}, this corresponds to a displacement in the free-stream of 27 diameters of the wind turbine, equal to the length of the wind tunnel test section; for this reason, the velocity realisations are assumed to be statistically independent.
Velocity snapshots have been phase-locked to the instantaneous position of the turbine rotor, to ensure a uniform distribution of all phases in the computed statistics; this has been realised by timing the laser discharge to the index signal of a rotary encoder installed on the turbine shaft.
A total of six phases have been recorded, at a distance of \SIlist{0;20;40;60;80;100}{\degree} from the reference rotor position; as the turbine rotor has three blades, the resulting velocity field is understood to be periodic to a \SI{120}{\degree} rotation of the turbine rotor.
A total of 300 instantaneous snapshots of velocity have been acquired for each phase: unless otherwise mentioned, statistics are computed on the full dataset consisting of 1800 snapshots per test case.
Uncertainty in the instantaneous velocity measurements has been estimated by DaVis to be equal to \SI{1.5}{\percent} of the measured values in the near wake fields of view and \SI{1.0}{\percent} in the far wake fields, with these values being constant between test cases.

\subsection{Test cases} \label{subsec:method-testcases}
The turbine wake has been generated under 18 different conditions: these are parametrised with the operating tip-speed ratio of the turbine and the freestream turbulence conditions that the turbine has been subject to.
The turbine has been operated at three distinct values of tip-speed ratio \tsr: these are $\tsr = 1.9$, for which the turbine generates low power and thrust, and the flow around the blades is mostly stalled; $\tsr = 3.8$, for which the turbine generates the most power; and $\tsr = 4.7$ at which the thrust generated is maximum.
Curves of power and thrust generated by the turbine have been previously reported by \cite{Gambuzza2021} as adimensional power and thrust coefficients (respectively \cp{} and \ct{}): these are defined as
    \begin{align}
        C_P &= \frac{Q \, \omega}{\frac{1}{2} \rho \, \uinf^3 \pi r_\mathrm{tip}^2}, \\
        C_T &= \frac{T}{\frac{1}{2} \rho \, \uinf^2 \pi r_\mathrm{tip}^2}, 
    \end{align}
where $Q$ is the torque generated by the rotor, $\omega$ is its angular velocity, and the product of these quantities is the mechanical power harvested by the turbine, $T$ is the turbine thrust, and $\rho$ is the air density.
The values of \ct{} generated by the turbine for the different \tsr{} at which it has been operated are listed in \cref{tab:ct-vs-tsr}: as reported in \cite{Gambuzza2021}, little effect of the freestream turbulence characteristics is seen on these values.
\begin{table}
    \centering
    \begin{tabular}{lr} \toprule
        \tsr{} & \ct{} \\ \midrule
        1.9 & 0.52 \\
        3.8 & 0.76 \\
        4.7 & 0.80 \\ \bottomrule
    \end{tabular}
    \caption{Turbine thrust coefficient \ct{} as a function of the tip-speed ratio \tsr{} for the operating conditions presented in this study.}
    \label{tab:ct-vs-tsr}

\end{table}

\begin{figure}[ht!]
    \centering
    \includegraphics{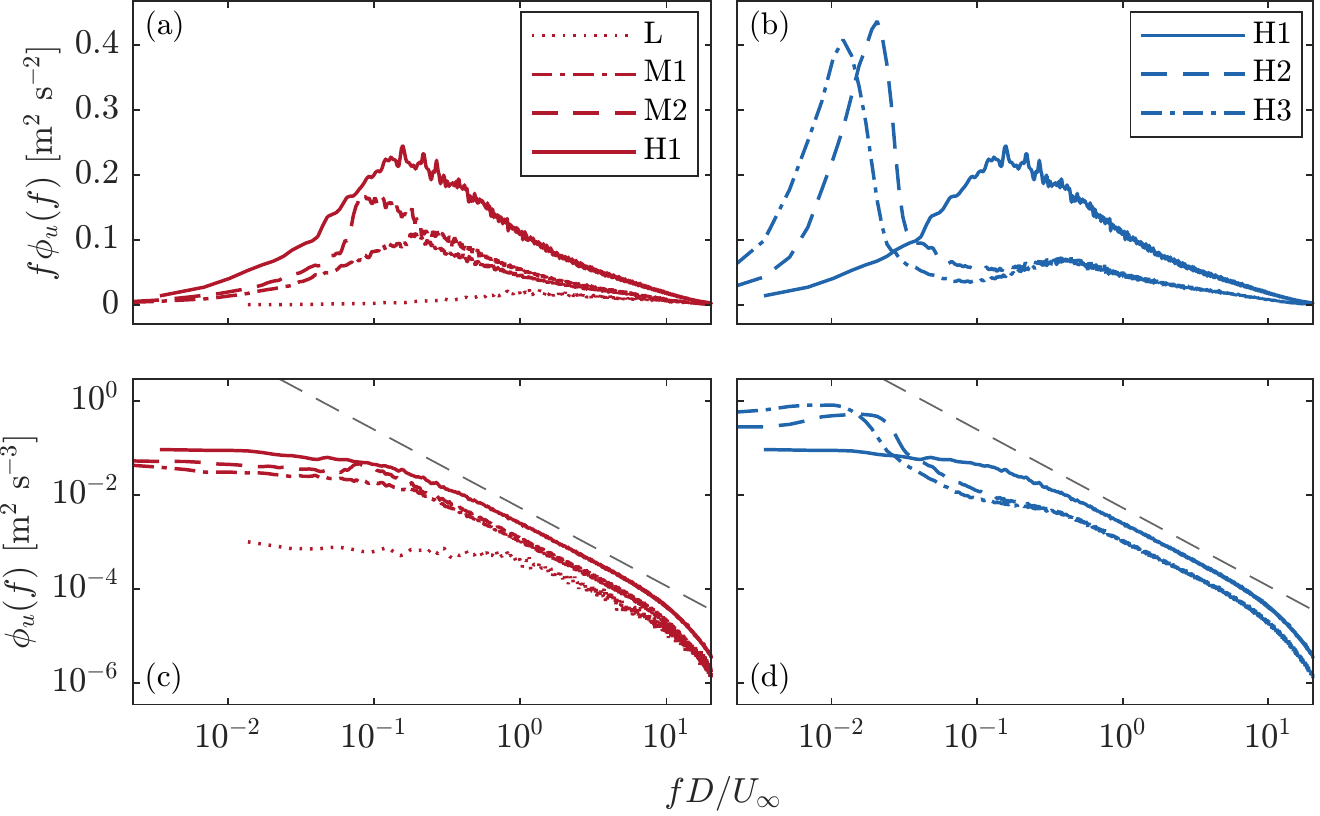}
    \caption{Top row: spectra of the streamwise component of freestream turbulence $\phi_u(f)$ premultiplied by the frequency axis \textit{versus} adimensional frequency for flows at integral time scale $\tscale{} \le 1$ (Kolmogorov-like flows, \textit{a}) and flows at constant turbulence intensity \iinf{} (non-Kolmogorov flows, \textit{b}). Bottom row: same spectra plotted as non-premultiplied on canonical log-log axes along with -5/3 slope (\textit{dashed grey}).}
    \label{fig:fst-spectra}
\end{figure}

The active grid has been used to generate six different freestream turbulence conditions: in this work, these are classified based on their turbulence intensity
\begin{equation}
	\iinf = \frac{\sqrt{\overline{u'^2}}}{\uinf},
\end{equation}
and their integral time-scale \tscale{}, computed as
\begin{equation}
	\tscale = \int_0^{\tau_0} \rho_{uu}(\tau) \, d\tau,
\end{equation}
where $\rho_{uu}(\tau)$ is the autocorrelation coefficient of $u'(t)$ and $\tau_0$ is the first value of $\tau$ for which $\rho_{uu}(\tau) = 0$.
This last quantity is presented in the remainder of the paper as normalised by the convective timescale $D/\uinf$.
To compute both \iinf{} and \tscale{}, hot-wire anemometry has been used to measure the freestream velocity in an otherwise empty test section, on a $5\times2$ grid at the centre of the test section spanning the rotor-swept area: more details on the hot-wire anemometry setup is included in \citet{Gambuzza2021}.
The characteristics of the six flows generated are summarised in \cref{tab:freestream-characteristics}.
\begin{table}
	\centering
	\begin{tabular}{lrrrrrr} \toprule
		Name                             & L   & M1  & M2  & H1   & H2   & H3   \\ \midrule   
		\uinf{} [\si{\meter\per\second}] & 8.0 & 8.0 & 8.0 & 8.0  & 8.0  & 8.0  \\ 
		\iinf{} [\si{\percent}]          & 3.0 & 7.5 & 8.8 & 11.5 & 11.3 & 11.6 \\
		\tscale{}                        & 0.1 & 0.9 & 1.0 & 1.0  & 6.3  & 10.5 \\ \bottomrule
	\end{tabular}
	\caption{Freestream turbulence characteristics of the generated inflow conditions.}
	\label{tab:freestream-characteristics}
\end{table}

The freestream turbulence spectra are presented in \cref{fig:fst-spectra}: in this, the flows are divided in two families, one exhibiting the canonical distribution of energy in the spectrum of \cite{Kolmogorov1941}, which are reported in \cref{fig:fst-spectra}(a, c) and are characterised by $\tscale \le 1$, and a second for which freestream turbulence intensity \iinf{} is approximately constant, but the distribution of this does not follow that of the Kolmogorov spectrum.
Flows H2 and H3 are thus labelled non-Kolmogorov-like flows in the remainder of the text.
This is a desired feature, as previous studies of wind turbine wakes in turbulence employ turbulence whose spectra is Kolmogorov-like \citep{Chamorro2009,Barlas2016,Deskos2020,Neunaber2021a}.
As it was discussed in \cref{sec:intro}, this is often an assumption that is not challenged in literature, namely that the distribution of power in the incoming turbulence spectrum does not affect the development of a wind turbine wake.
Note that, while flows H2 and H3 exhibit a large value of \tscale{}, it would be incorrect to assume those are the result of very-large-scale structures: assuming Taylor's hypothesis to hold true, an estimate for the large-eddy size can be given by 
\begin{equation}
    L_0 = \tscale{} \, D,
\end{equation}
which for flow H3 results in an approximate value of \SI{1.8}{\meter}, or twice the test-section width and three times the test section height.
It must therefore be accepted that, for flows H2 and H3 presented here, Taylor's hypothesis need not hold as the high integral time-scale of these flows is instead representative of a low-frequency change in the bulk freestream velocity as seen by the turbine.

\section{Results and discussion} \label{sec:results}
As briefly introduced in \cref{sec:intro}, one of the most important parameters of the turbine wake is the velocity deficit $\Delta{}U$.
This is defined as a function of the streamwise distance from the turbine as
\begin{equation}
    \frac{\Delta{}U}{\uinf} = \frac{\uinf - \min_y(U(x, y))}{\uinf},
\end{equation}
where the dependency of $U$ on $y$ is removed by taking the minimum in that direction.
This parameter is of interest to a wind farm designer as this sets the streamwise distance between rows of turbines in a wind farm.

\begin{figure}[ht!]
    \centering
    \includegraphics{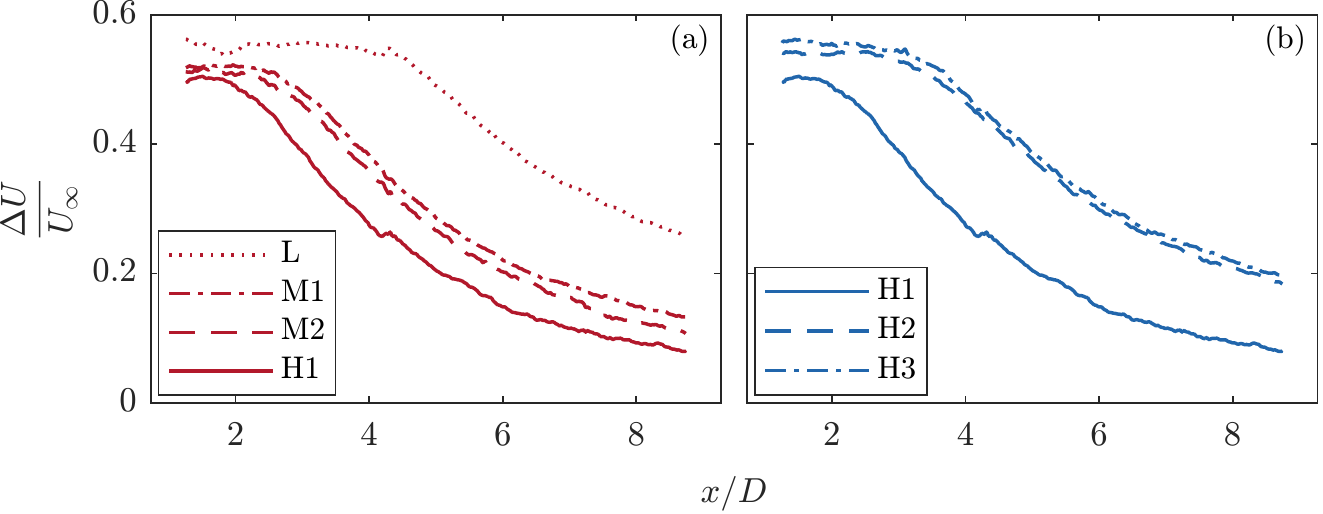}
    \caption{Maximum velocity deficit $\Delta{}U/\uinf$ as a function of the streamwise distance from the turbine $x/D$ for the Kolmogorov-like flows at $\tscale \le 1$ (\textit{a}) and for the constant \iinf{} flows (\textit{b}). Turbine operating at peak power-generating $\tsr = 3.8$.}
    \label{fig:velocity-deficit-tsr38}
\end{figure}
\Cref{fig:velocity-deficit-tsr38} reports the trends of the velocity deficit $\Delta U$ for all inflow conditions analysed in this study.
In particular, \cref{fig:velocity-deficit-tsr38}(a) isolates the trends for the Kolmogorov-like flows: it can clearly be seen that an increase in the freestream turbulence intensity generates a monotonic decrease of the velocity deficit, and therefore a faster wake recovery and a shorter overall wake length.
This is often connected to an increase in the turbulent mixing, favouring the homogenisation of velocity between the low-speed wake and the higher-momentum freestream surrounding it \citep{Medici2006,Chamorro2009}, a behaviour that is also seen for other bluff-bodies in turbulence \citep{Hearst2016a}.
On the other hand, \cref{fig:velocity-deficit-tsr38}(b) instead collects the family of flows at constant \iinf{}, and even in this case important differences between the wake velocity profiles are seen.
In particular, the wakes generated for the non-Kolmogorov-like flows H2 and H3 are seen to be only slightly different, and these appear to evolve more slowly both with respect to the equivalent-\iinf{} flow H1, and more surprisingly to flow M1, which has a much smaller value of freestream turbulence intensity.
In fact, only the wake generated under flow L, which here has been taken as reference for a low-turbulence flow, evolves more slowly.

\begin{figure}[ht!]
    \centering
    \includegraphics{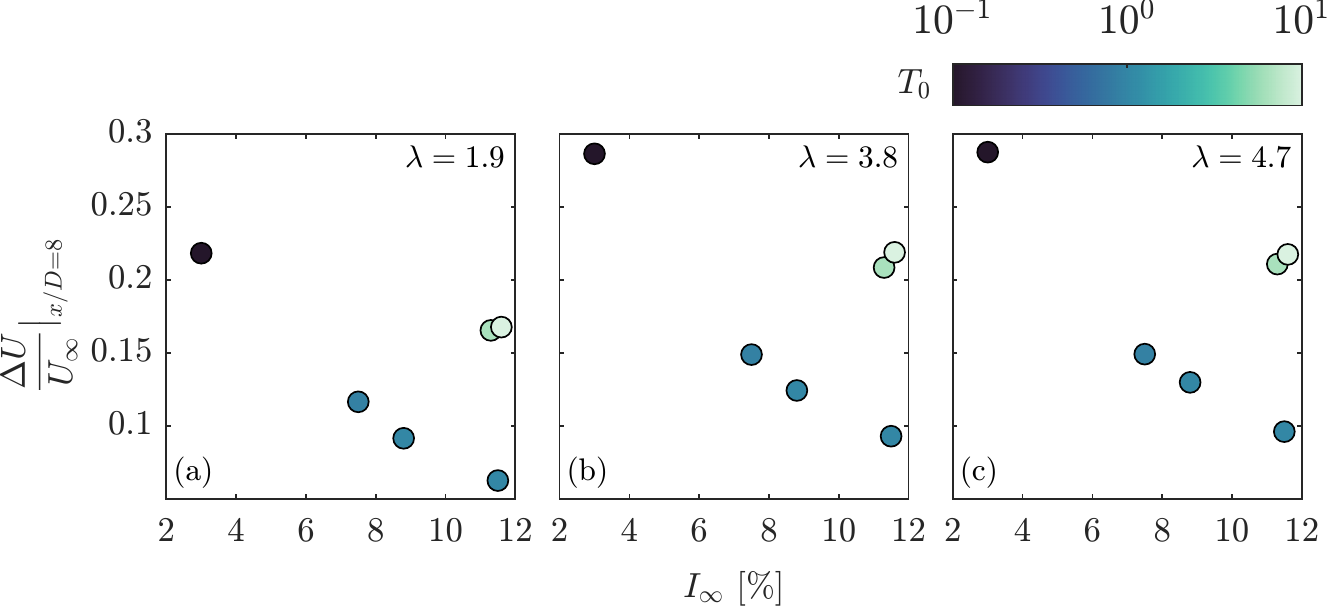}
    \caption{velocity deficit $\Delta{}U/\uinf{}$ at $x/D = 8$ for all operating conditions, as a fucntion of the freestream turbulence intensity \iinf{} (horizontal axis) and integral time scale \tscale{} (colour, note the logarithmic axis), for the turbine operating at $\tsr = 1.9$ (\textit{a}), $\tsr = 3.8$ (\textit{b}) and $\tsr = 4.7$ (\textit{c}).}
    \label{fig:velocity-deficit-all}
\end{figure}
Data reported in the previous \cref{fig:velocity-deficit-tsr38} has been obtained by fixing the turbine tip-speed ratio and, ultimately, its thrust. However, \cref{fig:velocity-deficit-all} reports the value of $\Delta{}U/\uinf$ for all flows the turbine has been subjected to and for all three values of \ct{} here studied; freestream turbulence intensity is reported on the horizontal axis and freestream integral time scale is reported as colour of the markers.
It can clearly be seen that the results of \cref{fig:velocity-deficit-tsr38} can be generalised to all values of \ct{} here studied, and thus these are representative of a peculiar behaviour of the wind turbine.
The trend of velocity deficit with turbulence intensity, limiting the analysis to the Kolmogorov-like flows of $\tscale \le 1$, is clearly decreasing; as no data has been acquired for values of $\iinf > \SI{12}{\percent}$, it cannot be said whether this tends asymptotically to zero or not.
This is an important distinction: an asymptotical trend to zero would suggest that the turbine wake can be arbitrarily shortened by a large enough value of freestream turbulence.
On the other hand, a trend to a small but finite value would instead hint to a behaviour for which the mean flow in the wake is unaffected by an increase of \iinf{} after a certain threshold, a behaviour that is similar to what observed for boundary layers generated in freestream turbulence \citep{Sharp2009,Dogan2016}.

\begin{figure}[ht!]
    \centering
    \includegraphics{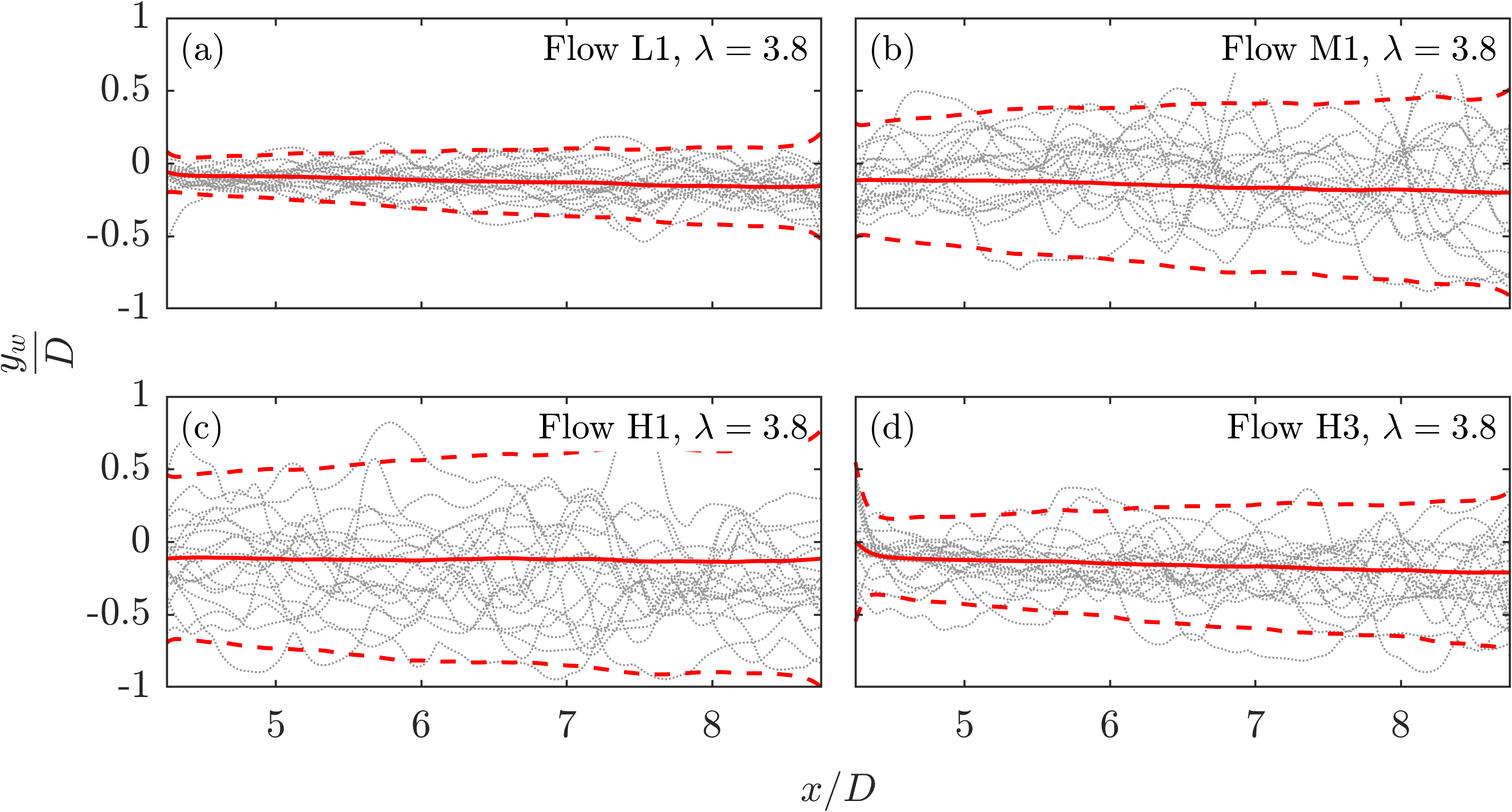}
    \caption{Instantaneous trajectory of the wake $y_w(x)$ (dotted grey lines, 20 random trajectories shown per test case), alongside mean wake trajectory (solid red line) and boundaries of the wake meandering region (dashed red lines). Data shown for inflow L (\textit{a}), M1 (\textit{b}), H1 (\textit{c}) and H3 (\textit{d}); turbine operating at $\tsr = 3.8$ for all four subfigures.}
    \label{fig:wake-meandering}
\end{figure}
Wakes of turbines are also seen to meander, that is change their instantaneous trajectory; this is a phenomenon that is understood to be driven both by instabilities in the shear layer and by the presence of large-scale eddies in the inflow \citep{Heisel2018}.
To identify the instantaneous trajectories of the wake from the individual velocity snapshots, the method from \citet{Howard2015a} is used: in this, the instantaneous wake trajectory is determined for each $x$ as
\begin{equation}
    y_w(x, t) = \underset{y}{\textrm{argmin}}(u(x, y, t)),
\end{equation}
which is then low-pass filtered to remove all contributions having a wavelength smaller than $D/2$.
\Cref{fig:wake-meandering} reports, for each of the edge cases at $\tsr = 3.8$, 20 randomly chosen instantaneous trajectories in dotted grey.
Alongside these, the mean wake trajectory is found by averaging the instantaneous $y_w$ in time, and the extent of the meandering region is reported as twice the standard deviation of the instantaneous wake trajectories.
In other words, this is the region in space where the trajectory of the turbine wake is located in \SI{95}{\percent} of the observations.
It can be appreciated that the width of this region increases both with distance from the turbine, a phenomenon observed for the wakes of solid and porous disks by \citet{Espana2011} and in LES of wind farms by \citet{Foti2019}, and with the freestream turbulence intensity content \iinf{}.
\begin{figure}[ht!]
    \centering
    \includegraphics{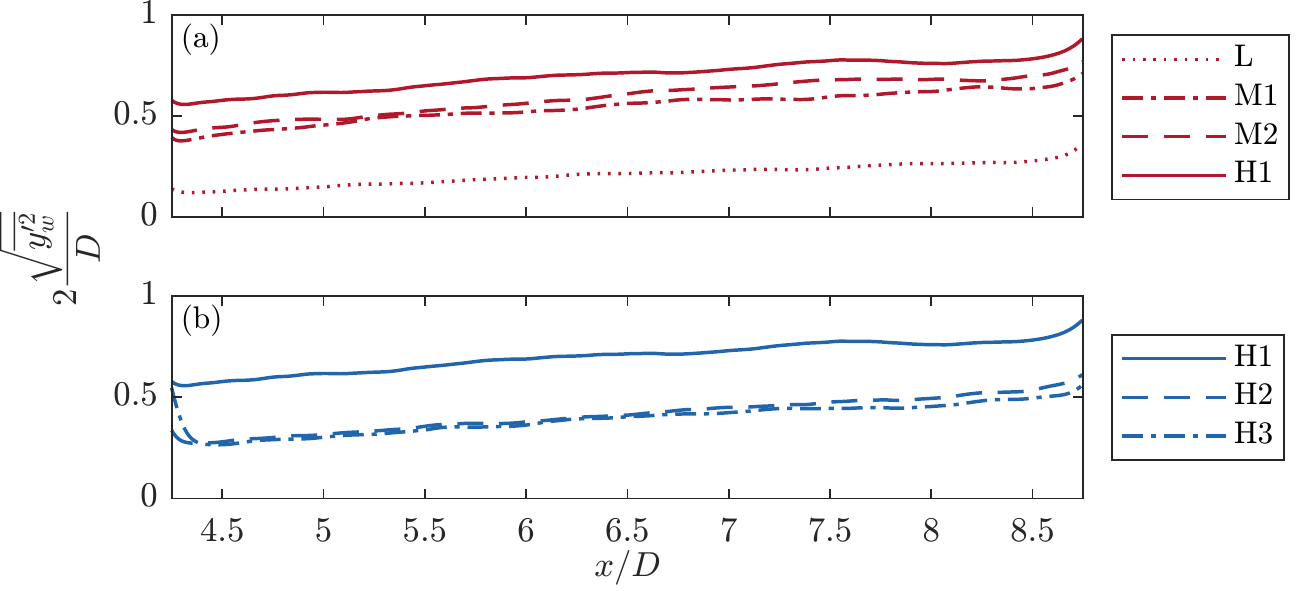}
    \caption{Extent of the meandering region estimated as twice the standard deviation of the instantaneous wake trajectories for the Kolmogorov-like flows at $\tscale \le 1$ (\textit{a}) and the flows at constant \iinf{} (\textit{b}). Turbine operating at $\tsr = 3.8$.}
    \label{fig:wake-meandering-extent}
\end{figure}
Once again, by limiting the analysis to the Kolmogorov-like flows and the extent of the wake meandering region (see data reported in \cref{fig:wake-meandering-extent}(a)), it is easy to appreciate that the extent of the wake meandering increases with the freestream turbulence content.
However, the largest increase happens between flows L and M1, in conjunction with the increase in integral time scale of the flow \tscale{}, suggesting that this is mostly due to the size of the eddies introduced in the freestream; adding more turbulence without affecting its scales increases the extent of meandering by inducing shear layer instabilities \citep{Medici2006}, although its effects are of smaller intensity when compared to those of mean-flow convection.
On the other hand, the non-Kolmogorov flows H2 and H3 are seen to result in little wake meandering whose extent is, like for the velocity deficit, lower than that generated by flow M1 but higher than that of the low-turbulence flow L.
This is despite a large value of \tscale{} that should suggest eddies 5 to 10 times larger between these flows and the equivalent-\iinf{} flow H1; however, one must remember the remark given in \cref{subsec:method-testcases}, for which Taylor's frozen hypothesis need not hold for the high-\tscale{} flows and the large value of \tscale{} is not necessarily representative of the large scale structure size.

\begin{figure}[ht!]
    \centering
    \includegraphics{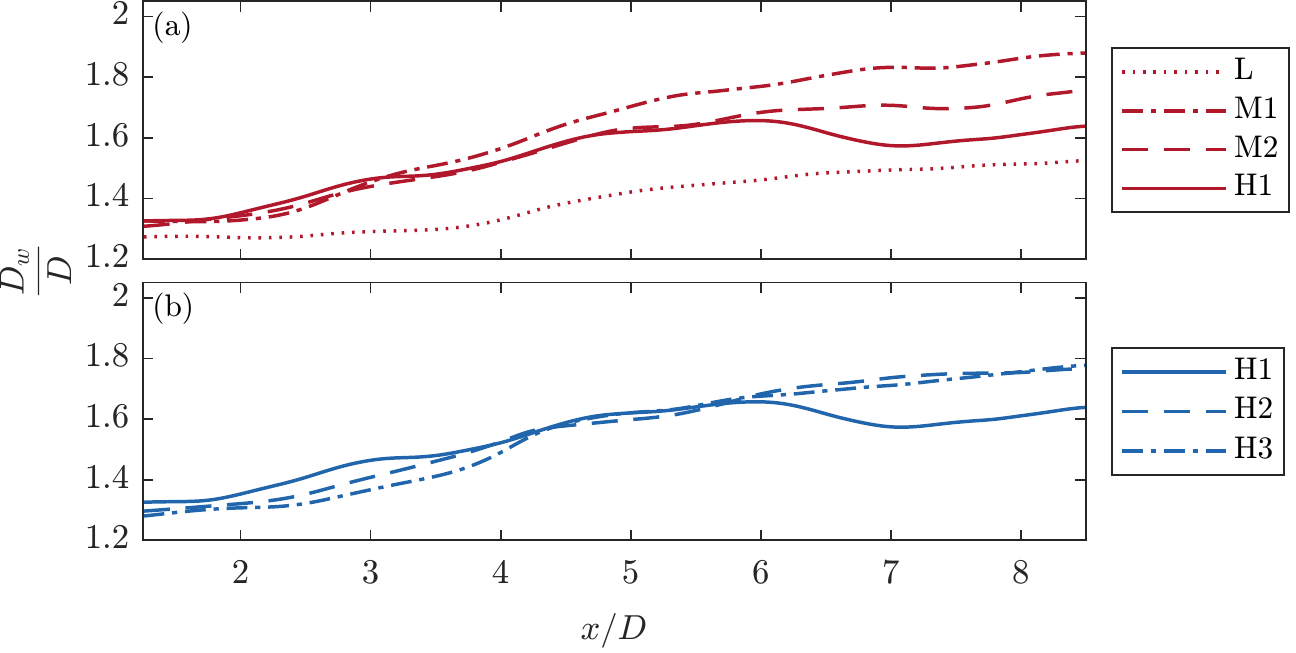}
    \caption{Adimensionalised wake diameter $D_w/D$ measured at $\tsr = 3.8$, for inflows with $\tscale \le 1$ (\textit{a}) and flows at constant \iinf{} (\textit{b}).}
    \label{fig:wake-diameter-tsr38}
\end{figure}
In addition to the velocity deficit, the wake diameter is an important parameter as this sets the lateral spacing between turbines in a farm; its growth in the streamwise direction is often taken as representative of the wake recovery rate and thus of the extent of the wake in the streamwise direction \citep{Jensen1983,Frandsen2006}.
\Cref{fig:wake-diameter-tsr38} reports the wake diameter $D_w$ measured for all investigated inflows with the turbine operating at $\tsr = 3.8$; as the presence of the mast affects the flow for $y/D < 0$, this is computed as twice the distance between the iso-line of $U = \uinf{}$ and the mean wake trajectory $\overline{y_w}$, which itself is obtained by averaging the instantaneous wake trajectories shown in the previous \cref{fig:wake-meandering}.
To highlight the large-scale trends, the trends of $D_w$ are also low-pass filtered to remove contributions with wavelengths below $D/2$.
The most immediate result that can be observed is that, for the investigated cases at $\tsr = 3.8$, little effect of the inflow is seen on the initial evolution of the wake and in particular of its slope: all flows except flow L result in wakes with similar diameters for $x/D < 4$.
This is an important observation, as this means that inferring the wake recovery rate $k$ from the wake diameter trend might lead to inaccurate estimations in the field.
For the high-turbulence flow H1, one can even observe that the trend of $D_w$ is not linear for all values of $x/D$, and instead plateaus after $x/D = 7$; while it could be argued that this is due to the presence of the wind tunnel walls, this does not seem to affect the evolution of all other wakes developed under different inflows, meaning the reason for this constant wake diameter might be found in the turbulence enveloping the wake.
Following the approach of \citet{Pope2000}, one can observe that the momentum deficit flows is an invariant of the wake equal to the turbine thrust.
The streamwise momentum equation can be written as
\begin{equation}
    T = \rho \, \left( \int_{A_i} \uinf^2 \, dA - \int_{A_w} U^2 \, dA \right),
    \label{eq:balance-momentum-zong}
\end{equation}
where $T$ is the turbine thrust, $\rho$ is the fluid density, and $A_i$ and $A_w$ are two sections of a stream tube containing the wind turbine far upstream and downstream of the turbine respectively.
As $T$ is constant, so must be the product $U^2 \, dA$ on $A_w$; however, it can be seen from the data presented in \cref{fig:velocity-deficit-tsr38}(a) and \cref{fig:wake-diameter-tsr38}(a) that for $x/D > 7$, $D_w$ and therefore $A_w$ is constant while $U$ changes.
It is at this point important to note that \cref{eq:balance-momentum-zong} only holds for $A_w$ sufficiently downstream of the turbine, so that the mean momentum convection $U\frac{\partial U}{\partial x}$ dominates over the turbulent transport in both the streamwise and stream-normal directions $\frac{\partial \overline{u'^2}}{\partial x}$ and $\frac{\partial \overline{u'v'}}{\partial y}$ respectively.
Data reported in \cref{subsec:results-farwake} will show that the case of flow H1 and $\tsr = 3.8$ is also the test case for which the terms of the Reynolds stress tensor are the largest; this, coupled with the fact that at $x/D \ge 7$, $U \simeq \uinf$, drives $U \frac{\partial U}{\partial x}$ to low values, makes the assumptions under which \cref{eq:balance-momentum-zong} not necessarily hold.

\begin{figure}[ht!]
    \centering
    \includegraphics{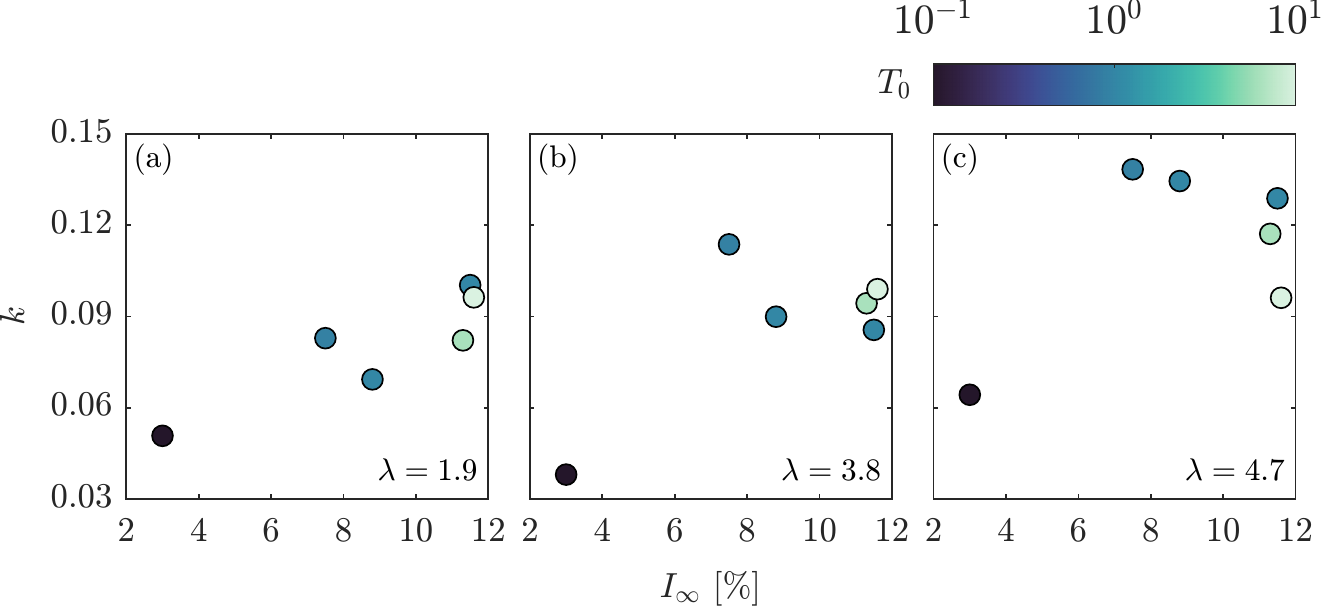}
    \caption{Wake growth rate $k$ found as the slope of the linear regression of $D_w/D$; data for $\tsr = 1.9$ (\textit{a}), $\tsr = 3.8$ (\textit{b}) and $\tsr = 4.7$ (\textit{c}). Data from subfigure (b) was already presented in \cref{fig:wake-diameter-tsr38}.}
    \label{fig:wake-growth-rate}
\end{figure}
The value of $k$ can be obtained, as a function of the freestream turbulence characteristics, as the slope of the linear regression of $D_w/D$; to account for the plateau in $D_w$ for the test case of flow H1 and $\tsr = 3.8$, the linear regression is performed for $x/D < 5$ for all test cases.
Data reported in \cref{fig:wake-growth-rate} shows that the trend of $k$ is erratic with both turbulence intensity and integral time scale: for the low thrust case of $\tsr = 1.9$ (\cref{fig:wake-growth-rate}(a)) a somewhat linear trend of $k$ with \iinf{} is seen, with a small effect of \tscale{} on the wake expansion; this last observation is similarly seen for $\tsr = 3.8$, for which however the trend of $k$ with \iinf{} is not linear, and the wake that experiences the largest expansion is that developed under the moderate-turbulence of flow M1.

The data presented in this section therefore shows how the wake generated by the isolated turbine is affected by both the freestream turbulence intensity \iinf{} and its frequency content, which in this study has been parametrised as its integral time scale \tscale{}.
However, the approach followed in this section has not attempted to quantify these effects, nor it has explained the physical phenomena that drive this evolution or these trends.
The next section will therefore present a parametrisation of the wake generated by the wind turbine based on frequently used analytical wake models, which will then be leveraged in the next sections to quantify the effects of turbulence on the wake developed by the turbine.

\subsection{Quantifying the effects of turbulence} \label{subsec:results-models}
As mentioned in \cref{sec:intro}, analytical wake models are powerful tools that are often used both in literature and in the field to represent the complex turbine wake in a simplified fashion.
These operate by assuming the existence of a relationship between the velocity in any point of the wake and a reduced set of parameters: for instance, the widespread Gaussian wake model, developed by \cite{Bastankhah2014}, assumes that the velocity in any point of the wake is described by the set of equations
    \begin{gather}
        \frac{\Delta{}U}{\uinf} = \left( 1 - \sqrt{1 - \frac{\ct{}}{8 (\sigma_w/D)^2}} \right) \mathrm{exp}\left(-\frac{(r/D)^2}{2(\sigma_w/D)^2}\right), \label{eq:bastankhah-1} \\
        \frac{\sigma_w}{D} = k^* \frac{x}{D} + \epsilon, \label{eq:bastankhah-2}
    \end{gather}
where $r$ is the radial coordinate away from the turbine axis, $k^*$ is a parameter that serves the role of the wake expansion rate $k$ in the previous section, and
    \begin{align}
        \epsilon &= 0.2 \sqrt{\beta}, \label{eq:definition-epsilon}\\
        \beta{} &= \frac{1}{2} \frac{1 + \sqrt{1-\ct}}{\sqrt{1-\ct}}.
    \end{align}
Regarding \cref{eq:definition-epsilon}, the authors state that $\epsilon$ should be equal to $0.25 \sqrt{\beta}$ from considerations on the total mass flow deficit rate at $x/D = 0$; however, this is often used as presented in \cref{eq:definition-epsilon} as it yields estimates that better match with measurements.

\begin{figure}[ht!]
    \centering
    \includegraphics{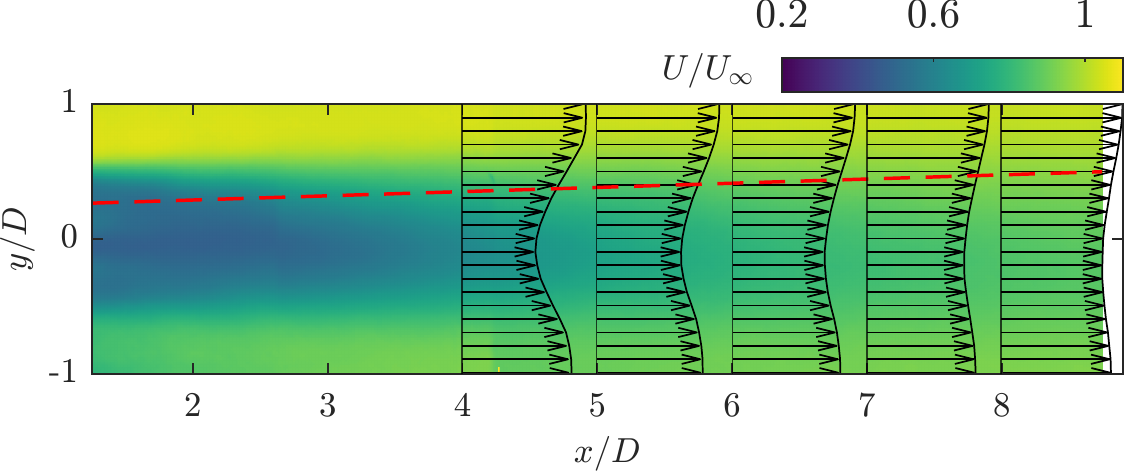}
    \caption{Determination of $k^*$ from PIV measurements in the mean wake. In colour: mean streamwise velocity component; in black arrows: selected velocity profiles; in dashed red: linear regression of $\sigma_w(x)$. Data for flow M1, $\tsr = 3.8$.}
    \label{fig:kstar-pivfield}
\end{figure}
As full knowledge of the wake geometry is known from the PIV velocity fields, one can compute the value of $k^*$ for each operating condition and inflow as outlined in \cref{fig:kstar-pivfield}: for every value of $x/D$, a velocity profile is obtained and fitted to a Gaussian profile, with the standard deviation of the fitted profile being an estimate of $\sigma_w(x)$.
The value of $k^*$ is then the slope of the linear regression of this last quantity.
As for the previous data on the wake diameter $D_w$ presented in \cref{fig:wake-diameter-tsr38}, the fitting is limited to the range $0 < y/D < 1$ as the presence of the mast has impacted the symmetry of the wake; moreover, as the velocity profiles need not be Gaussian close to the turbine, especially at low values of \iinf{} \citep{Medici2006}, $\sigma_w$ is only computed for $x/D \ge 4$.

\begin{figure}[ht!]
    \centering
    \includegraphics{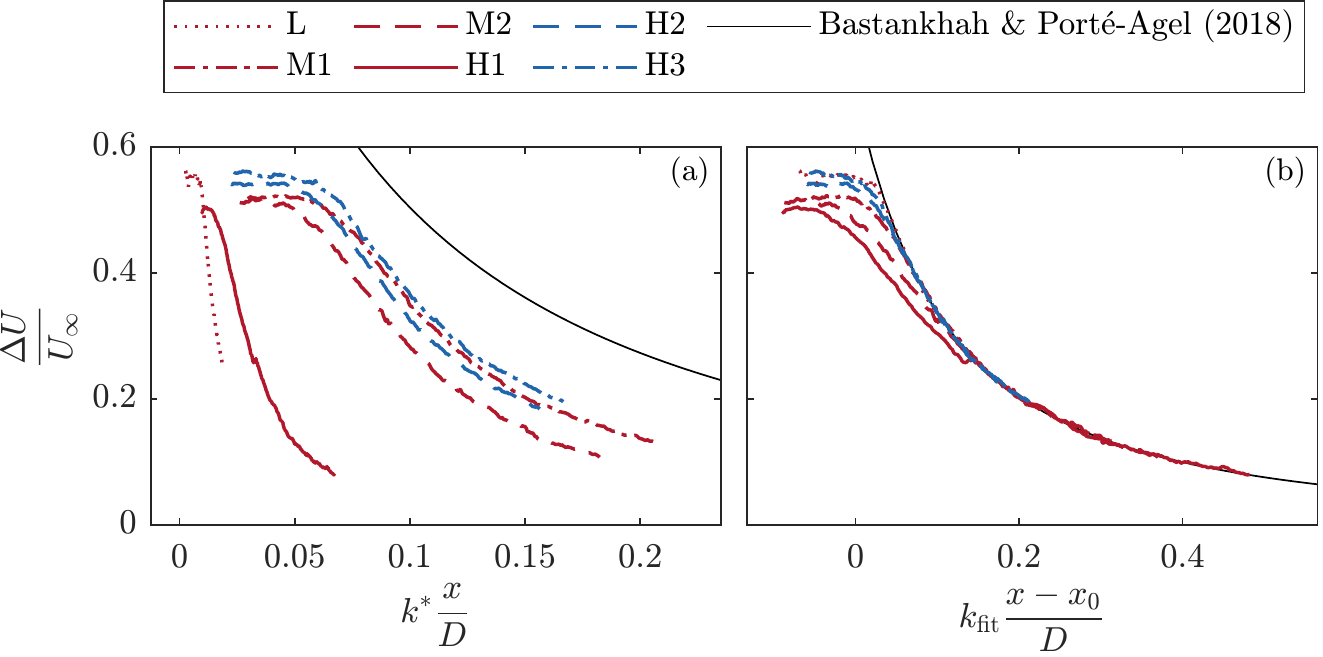}
    \caption{Comparison of the wake velocity deficit trends $\Delta{}U/\uinf$ with the trend predicted by \cite{Bastankhah2014}, assuming $\epsilon_0 = 0.2 \sqrt{\beta}$ and $k^*$ inferred from wake profile fitting (\textit{a}), and with the trends obtained by determining the best-fit values of $k^*$ and $x_0$ assuming $\epsilon_0 = 0.25 \sqrt{\beta}$ (\textit{b}). Data for $\tsr = 3.8$.}
    \label{fig:wake-fitting}
\end{figure}
\Cref{fig:wake-fitting}(a) reports, for all wakes obtained at $\tsr = 3.8$ and thus constant $\ct = 0.76$, the trends of velocity deficit previously reported in \cref{fig:velocity-deficit-tsr38} where the values on the horizontal axis are the distance from the turbine multiplied by the $k^*$ attained in each specific operating condition: it can be appreciated that none of the curves collapse on the prediction that the \cite{Bastankhah2014} offers.
Instead, \cref{fig:wake-fitting}(b) reports the same trends as in \cref{fig:wake-fitting}(a) having normalised the distance from the wind turbine by using two parameters: a virtual origin $x_0$ and an alternative value of $k^*$, which has been labelled \kfit{}.
Both the virtual origin $x_0$ and the alternative wake recovery rate \kfit{} have been obtained as the two parameters that minimise the sum of squared residuals between each measured trend of $\Delta{}U$ and the predicted curve according to \cref{eq:bastankhah-1}, having imposed $r/D = 0$; moreover, it has also been imposed that $\epsilon = 0.25 \sqrt{\beta}$.
Note that the use of a virtual origin to describe the evolution of bluff body wakes in turbulence is a widely employed method for other bluff bodies such as porous disks \citep{Rind2012,Rind2012a,Pal2015}, spheres \citep{Spedding1996}, or even when describing wakes of bodies in turbulent boundary layers \citep{Sakamoto1983}; this approach has been recently shown to be valid for wakes generated by utility-scale turbines in the atmospheric boundary layer \citep{Neunaber2022}.
While the physical meaning of the wake recovery rate is immediate, the plotted trends in \cref{fig:wake-fitting}(b) highlight how the virtual origin $x_0$ is loosely connected to the extent of the interval of $x$ for which the velocity deficit is constant.
In this case it can be seen that a collapse, especially for large values of $x/D$, is obtained regardless of the inflow conditions.
While this shows that the Gaussian wake model can be tuned to predict the wakes of turbines in a wide array of both realistic and non-Kolmogorov-like flows, this also allows to quantify the effect of turbulence on the wake generated by a wind turbine.

\begin{figure}[ht!]
    \centering
    \includegraphics{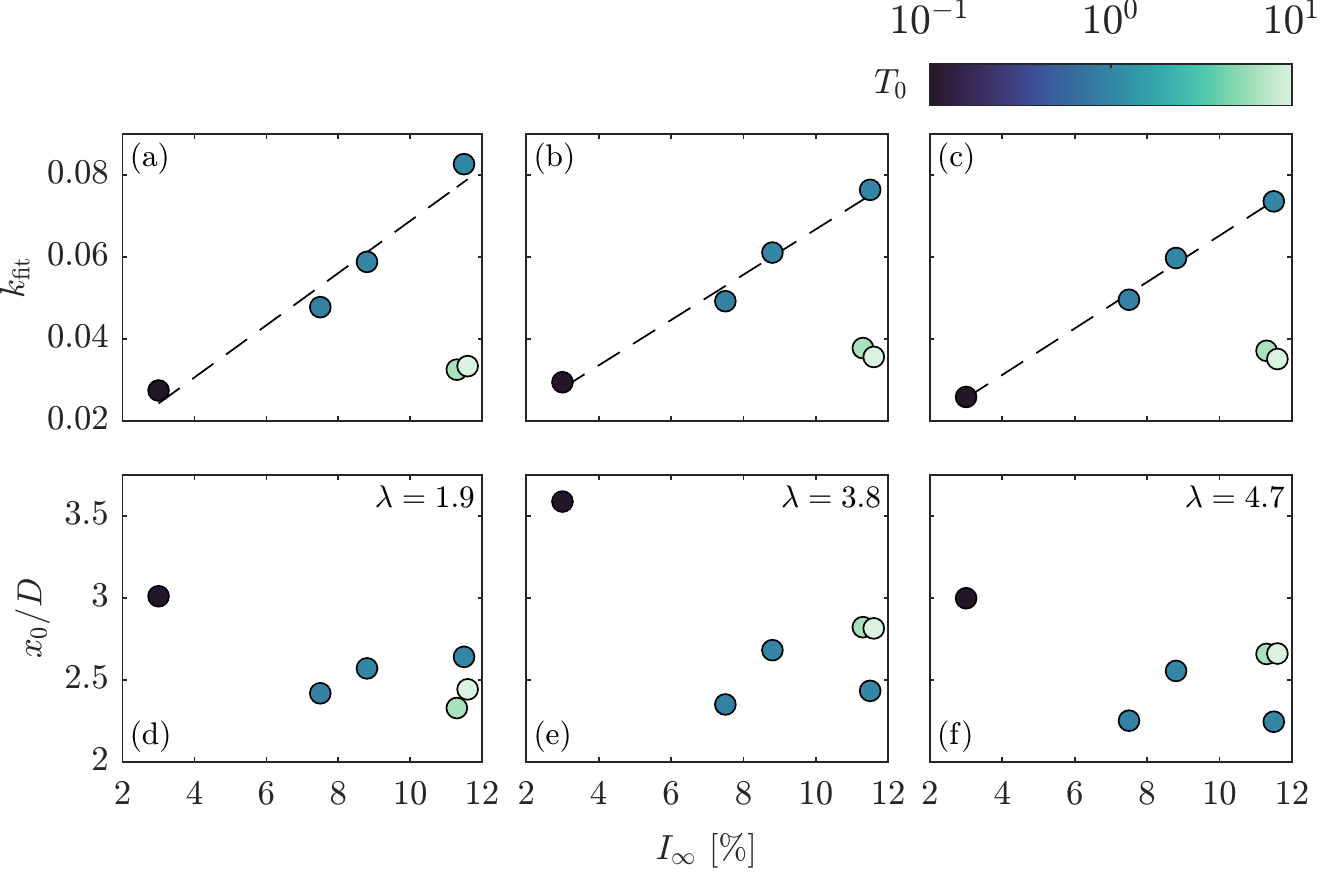}
    \caption{Trends of \kfit{} (\textit{a}, \textit{b}, \textit{c}) and $x_0/D$ (\textit{d}, \textit{e}, \textit{f}) with the freestream turbulence properties \iinf{} (horizontal axis) and \tscale{} (colour axis). Data for $\tsr = 1.9$ (\textit{a}, \textit{d}), $\tsr = 3.8$ (\textit{b}, \textit{e}), and $\tsr = 4.7$ (\textit{c}, \textit{f}).}
    \label{fig:wake-fitting-parameters}
\end{figure}
\Cref{fig:wake-fitting-parameters} reports the values of the two parameters used, \kfit{} and $x_0$, for all test cases investigated in this study.
For all of the turbine operating points, it can be seen that \kfit{} assumes a linear trend with \iinf{} if one limits the analysis to the Kolmogorov-like flows.
This is in good agreement with previous literature, namely the work of \citet{Niayifar2016} and that of \citet{Pena2016} on the Jensen wake model \citep{Jensen1983}, where the authors show how the linear relationship between the wake recovery rate and \iinf{} can be recovered assuming the vertical velocity profile is well described by Monin-Obukhov similarity theory \citep{Monin1954}; one must however note that no vertical velocity profile is present in our measurements, and thus the linear relationship between \kfit{} and \iinf{} holds even outside the atmospheric boundary layer.
The general trend of \kfit{} mimics well what already seen on the wake velocity deficit, for instance in \cref{fig:velocity-deficit-tsr38} and \cref{fig:velocity-deficit-all}, where wakes exhibiting higher values of $\Delta{}U$ far from the turbine have low values of \kfit{} and vice-versa.

Instead, the virtual origin $x_0$ is seen to be less affected by the freestream turbulence conditions: generally, this value is the highest for the low-turbulence case of flow L, regardless of the turbine thrust coefficient, and all other flows exhibit a value of $x_0/D \simeq 2.5$, only moderately affected by \ct{} and free-stream turbulence.
It is also interesting to note that, limiting our attention to high-turbulence flows, the value of $x_0$ is higher for the non-Kolmogorov flows for the two test cases at high \ct{} than it is for the Kolmogorov-like flow H1, while this trend appears reversed for the low-\ct{} test case of $\tsr = 1.9$.

The parametrisation of the wake in its recovery rate \kfit{} and its virtual origin $x_0$ also provides a convenient separation between two regions in the wake: in the next section we will argue that the extent of the near wake is loosely related to the trends of $x_0$ and to the stability of the helical vortex set enveloping the wake.
Conversely, one can interpret the value of \kfit{} to be the main parameter that drives the wake evolution for large distances downstream of the turbine, and thus to be representative of the far wake evolution.

\subsection{Influence on near wake extent} \label{subsec:results-nearwake}
Customarily, the wake of a wind turbine is often divided in two regions: a near wake, where the rotor geometry affects the local velocity field, and a far wake which is instead independent of the turbine geometry and self-similarity is attained \citep{DeCillis2020}; however, seldom a quantitative definition of their extent is given.
The division of the wake in a near and far field is commonplace for wakes generated by all bluff bodies and goes back to the first works by \citet{Castro1971} and \citet{Pope1976}.
Attempts to define the location of the boundary between the near and the far wake have been, in recent times, published by a number of authors: for instance, \citet{Sorensen2015} defines the boundary as the point of inflection of the turbulent kinetic energy content of some selected POD modes in the turbine wake; \citet{DeCillis2020} instead defines the boundary to be the location for which the time-averaged turbulent kinetic energy in the wake falls below a given threshold; \citet{Wu2012} uses instead the change in sign in the turbulent kinetic energy advection term.
In any case, one must note that the transition between near and far wake is qualitative in all cases, and no hard boundary between these two regions is present: therefore, any estimation of this parameter cannot but provide a qualitative information, and not a quantitative one.
This however does not mean there is no merit in estimating the near wake length, as trends of these estimations can still provide valuable information as comparison between test cases.
In this work, the definition of near-wake length is slightly changed from the one presented by \citet{Wu2012}: having defined the turbulent kinetic energy as
\begin{equation}
    \kappa = \frac{1}{2} \left( \overline{u'^2} + \overline{v'^2} + \overline{w'^2} \right),
    \label{eq:k-definition}
\end{equation}
its mean advection is
\begin{equation}
    \vec{U} \cdot \nabla\kappa = U\frac{\partial \kappa}{\partial x} + V\frac{\partial \kappa}{\partial y} + W \frac{\partial \kappa}{\partial z},
    \label{eq:k-adv}
\end{equation}
where $\vec{U} = (U; V; W)$ is the time-averaged velocity vector.
Note that for steady flows, this is equal to the material derivative of $\kappa$:
\begin{equation}
    \frac{D \kappa}{Dt} = \frac{\partial \kappa}{\partial t} + \vec{U} \cdot \nabla\kappa = \vec{U} \cdot \nabla\kappa,
\end{equation}
as $\frac{\partial}{\partial t} = 0$ in steady-state conditions.
The near wake is then defined as the location where $\frac{D\kappa}{Dt} > 0$ and vice-versa for the far wake.
Physically, the interpretations is as follows: as a flow parcel traverses the rotor-swept plane, its turbulent kinetic energy increases under the effect of both the circulation generated by the blades and the vorticity these shed, either as a vortex sheet or as the system of tip- and root-vortices.
As the parcel travels downstream, it will cede this to its surrounding; therefore, a positive value of this material derivative denotes that the evolution of the particle is driven by its interaction with the turbine and vice-versa.

As the data collected consists of the two in-plane components of velocity $u$ and $v$, it is impossible to compute the full turbulent kinetic energy as defined in \cref{eq:k-definition} as no information is available on $w'$; similarly, without $W$, the term relative to the advection of $\kappa$ in the $z$-direction cannot be computed.
To this reason, the definitions in \cref{eq:k-adv,eq:k-definition} have been implemented by assuming that $\frac{v}{w} = O(1)$ and $\frac{\partial{}}{\partial{}y} \simeq{} \frac{\partial{}}{\partial{}z}$; thus, the turbulent kinetic energy is estimated as
\begin{equation}
    \kappa = \frac{1}{2} \left(\overline{u'^2} + 2 \overline{v'^2} \right),
\end{equation}
and its material derivative is estimated as
\begin{equation}
    \frac{D\kappa}{Dt} =
      U \frac{\partial\kappa}{\partial x} + 
    2 V \frac{\partial\kappa}{\partial y},
\end{equation}
which is an assumption that holds true particularly close to the wake centre, as the wake is somewhat axisymmetric, and gets progressively less so farther from that, as an azimuthal velocity component is introduced as resultant of the torque generated by the blades and the assumption that $v \simeq w$ need not hold \citep{Medici2006}.

\begin{figure}[ht!]
    \centering
    \includegraphics{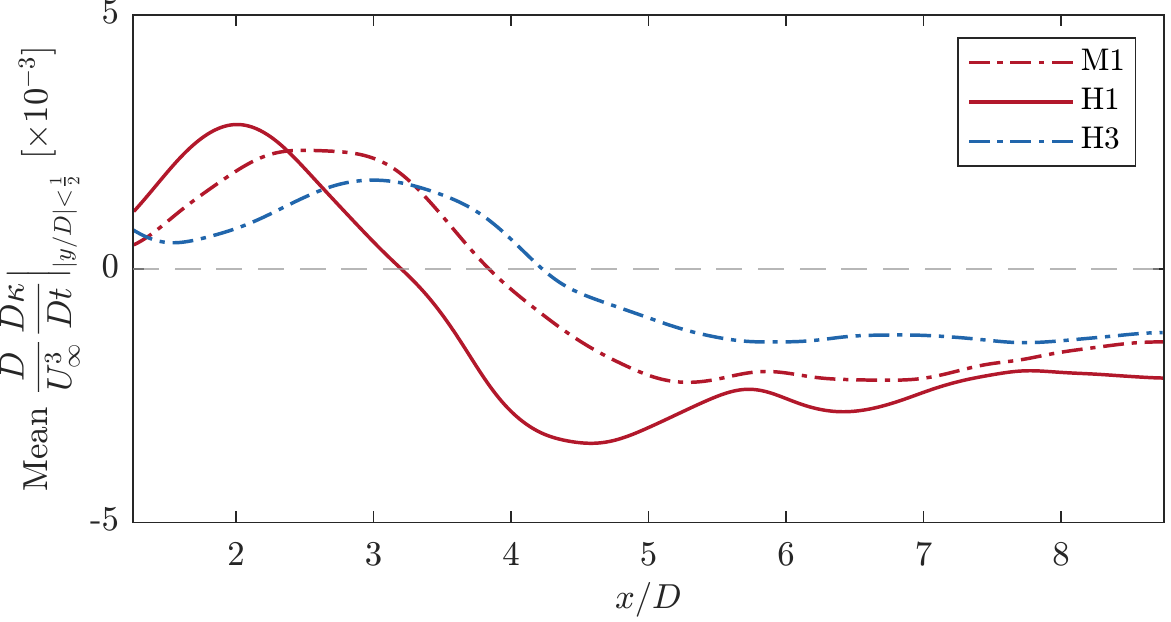}
    \caption{Trends of mean material derivative of turbulent kinetic energy $\frac{D\kappa}{Dt}$ for $|y/D| < \frac{1}{2}$ as a function of the streamwise distance from the turbine for inflows M1, H1, and H3. Turbine operating at $\tsr = 3.8$.}
    \label{fig:tke-trend}
\end{figure}
\begin{figure}[ht!]
    \centering
    \includegraphics{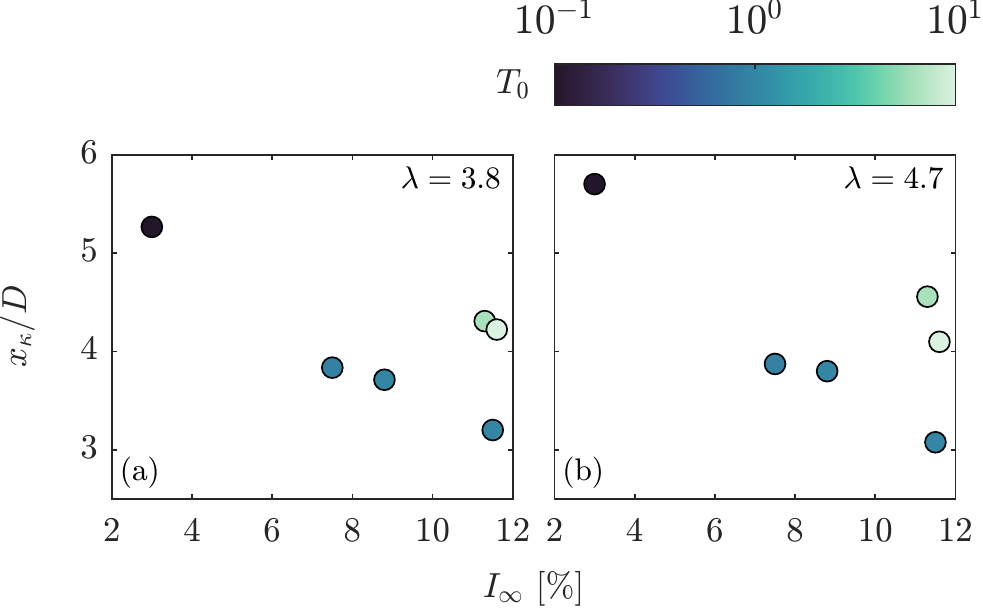}
    \caption{Location of the change of sign $x_\kappa/D$ in the spatial mean of ${D\kappa}/{Dt}$ as a function of the freestream turbulence intensity \iinf{} (horizontal axis) and integral time scale \tscale{} (colour axis), shown for $\tsr = 3.8$ (\textit{a}) and $\tsr = 4.7$ (\textit{b}).}
    \label{fig:x-kappa}
\end{figure}

\Cref{fig:tke-trend} reports the trends of the material derivative of $\kappa$ with distance from the turbine; this is adimensionalised by the term $\frac{D}{U_\infty^3}$.
To reduce the effect of experimental noise on the measurements, the trends reported have been obtained by averaging the value of $\kappa$ for $|y/D| < 1/2$, equivalent to limiting this analysis only to particles that have traversed the turbine rotor.
Moreover, each trend is low-pass filtered to remove all components having a wavelength smaller than $D/2$.
The trends highlight a clear distinction between the two regions at positive and negative $\frac{D\kappa}{Dt}$, as well as showing that this value does not trend to a zero value at long distances from the turbine.
This is expected and is due to the natural decay of turbulence with freestream distance, proper of both grid-generated turbulence \citep{Kistler1966,Hearst2016} and bluff-body wakes \citep{Wygnanski1986}.
Defining the value of $x$ at which $\frac{D\kappa}{Dt}$ changes sign as $x_\kappa$, one can observe the trends of this parameter with freestream turbulence characteristics in \cref{fig:x-kappa}.
Data for $\tsr = 1.7$ is not reported as $\frac{D\kappa}{Dt}$ is negative along the whole domain and no change in sign is observed; for this operating condition, the torque generated by the blades is marginally lower than that at $\tsr = 4.7$ and considerably lower than that at $\tsr = 3.8$ \citep{Gambuzza2021}: the circulation generated by the blades is therefore lower and so is the intensity of the tip- and root-vortices generated by the blades.
This, paired with a lower \tsr{} and therefore more spaced vortices, might have led to a less intense interaction between the turbine and the flow traversing the rotor-swept plane.
From the data here reported, it is immediate to notice the trends of $x_\kappa/D$ resemble well those of the virtual origin $x_0/D$ presented previously in \cref{fig:wake-fitting-parameters}(\textit{d--f}): both $x_0$ and the near wake length are reduced by an increase in freestream turbulence intensity \iinf{} and increased with an increase of the integral time scale \tscale{}.
However, data for $x_\kappa$ is not matching $x_0$ in magnitude; while the virtual origin is, with good approximation, equal to $2.5 D$ for all flows bar the low-turbulence conditions of test case L, $x_\kappa$ ranges instead from $3 D$ to $4.5 D$, with more visible differences between test cases.

\begin{figure}[ht!]
    \centering
    \includegraphics{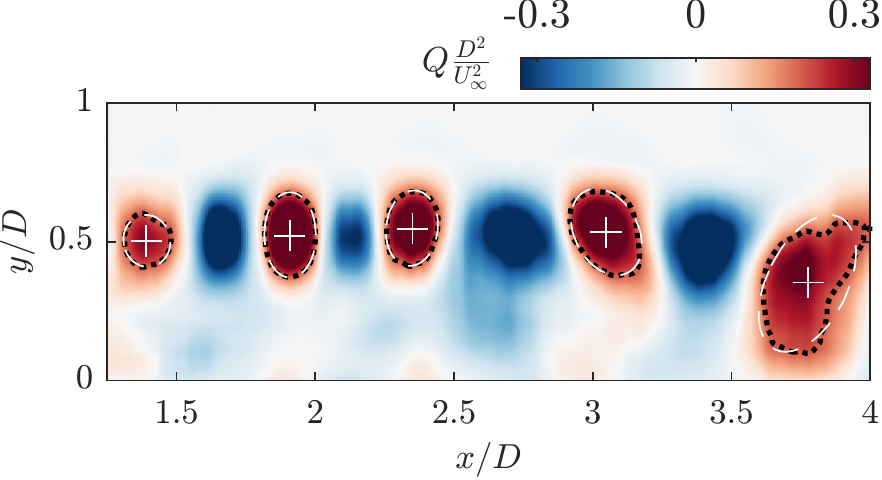}
    \caption{Identification of the tip-vortices location (\textit{white plus signs}) from isocontours of $Q$-criterion (\textit{black dotted}) approximated with best-fitting ellipses (\textit{white dashed}).x Data displayed is that of a representative snapshot obtained at $\tsr = 3.8$ and inflow L.}
    \label{fig:vortices-instantaneous}
\end{figure}
To understand the conditions that drive the evolution of the near wake, it is useful to observe the behaviour of the tip-vortices shed by the turbine.
Vortex identification from an instantaneous velocity snapshot can be carried out with a number of techniques; in this work, the $Q$-criterion \citep{Hunt1988,Jeong1995,Haller2005} has been employed due to the simplicity of the underlying equations and the ease of adaptation to planar data.
For an instantaneous velocity field $\vec{u}(t) = (u(t), v(t))$, the in-plane value of $Q$ is computed as
\begin{equation}
    Q = -\frac{1}{2} \left(\left(\frac{\partial u}{\partial x}\right)^2 + 2 \frac{\partial u}{\partial y} \frac{\partial v}{\partial x} + \left(\frac{\partial v}{\partial y}\right)^2\right),
\end{equation}
and vortices are individuated as continuous regions in the flow where $Q>0$.
\Cref{fig:vortices-instantaneous} reports the map of $Q$ for a representative velocity snapshot obtained for the operating conditions of $\tsr = 3.8$ and inflow L, along with the methodology used to individuate the location and intensity of individual vortices.
Given a computed map of $Q$, contiguous regions of positive $Q$ are individuated as the regions bounded by iso-lines of $Q$ equal to a certain threshold, which in this case has been chosen as $0.3 (D/\uinf)^2$; the iso-lines are then approximated with the best-fitting ellipse, with the centre of the ellipse being used as the location of the vortex and the value of $Q$ at the centre being an indication of the vortex intensity.

\begin{figure}[ht!]
    \centering
    \includegraphics{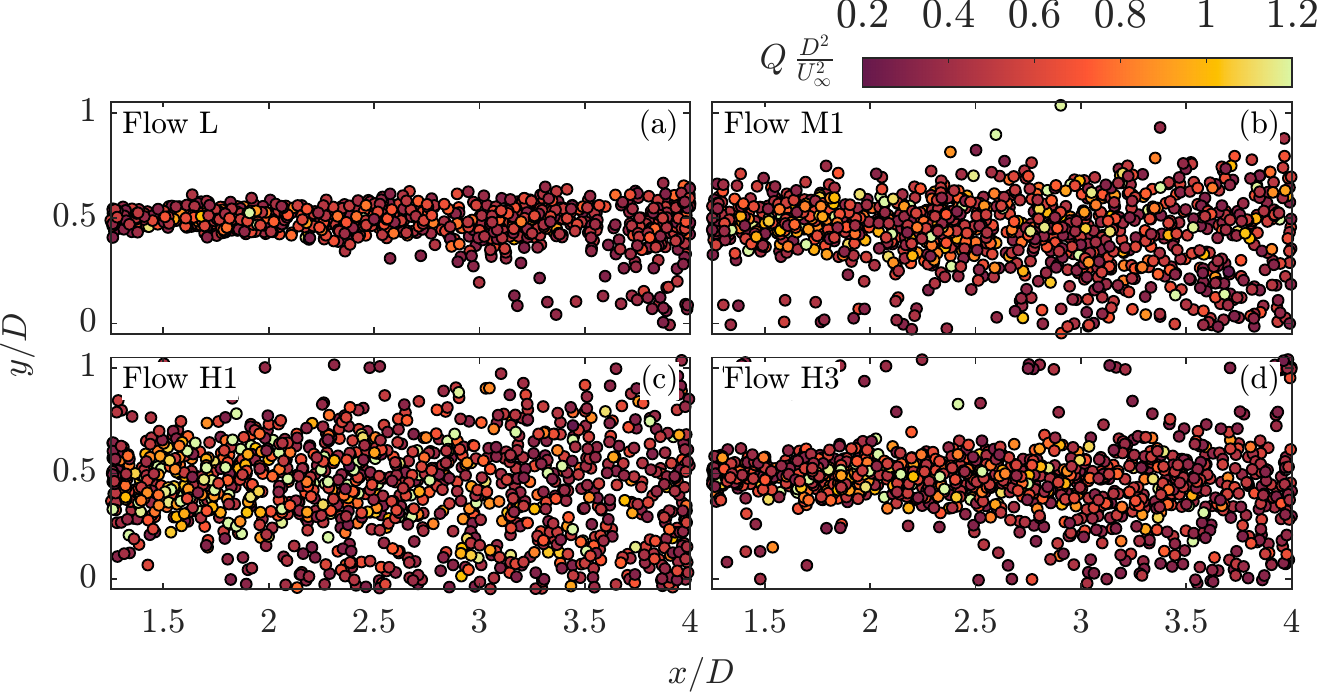}
    \caption{Instantaneous position of the tip-vortices for inflows L (\textit{a}), M1 (\textit{b}), H1 (\textit{c}), and H3 (\textit{d}), coloured by their peak $Q$. Turbine operating at $\tsr = 3.8$. For clarity, a random subset of 1200 individual vortices is shown for every subfigure.}
    \label{fig:vortices-positions}
\end{figure}
The position of the vortices is shown in \cref{fig:vortices-positions} for four edge-cases, all having the turbine operating at $\tsr = 3.8$.
From these, it can be seen how the low turbulence intensity of flow L results in a very coherent and stable train of vortices mostly concentrated around the tip-height line at $y/D = 0.5$, with some small instability setting on at $x/D > 3.5$.
For the two Kolmogorov-like inflows at high \iinf{}, namely M1 and H1, shear layer instability is favoured by the high amount of freestream turbulence and the vortices positions are more erratic: this shear layer instability in turn drives the wake meandering seen in \cref{fig:wake-meandering} for the same two test cases.
The test case of the non-Kolmogorov flow H3 is instead seen, despite the same value of \iinf{}, to result in an initially more stable shear layer, with the tip-vortices occupying a region close to $y/D = 0.5$, larger than that generated under inflow L but visibly bounded, unlike that of the two previous flows M1 and H1; this initial shear layer stability is possibly connected to the less intense wake meandering observed in \cref{fig:wake-meandering}.

It has been previously reported in literature that the presence of the helical vortex set envelopes the wake and prevents transfer of momentum between the wake and the surrounding freestream, thus delaying the wake evolution \citep{Lignarolo2014,Lignarolo2015,DeCillis2020}: this suggests that the initial wake evolution is hampered by the presence of a stable shear layer and thus by a train of coherent vortices, which effectively shield the near wake from the freestream.
This can be easily visualised, for the test cases here reported, with quadrant analysis to characterise the events on the mean vortices trajectory.
At these locations, a $v'<0$ denotes flow crossing the vortices trajectory from the high-momentum freestream to the wake and $v'>0$ denotes conversely motion from the wake outwards; simultaneously, $u'>0$ indicates motion of high-momentum flow and $u'<0$ indicates that of low-momentum flow.
Wake evolution is therefore promoted either by the sweep of high-momentum flow inside the wake, for which $u'>0$ and $v'<0$, or by ejection of low-momentum flow from the wake, for which $u'<0$ and $v'>0$; vice-versa, wake evolution is hampered by interaction events, for which $u'$ and $v'$ have the same sign.
Defining thus the individual contributions to the in-plane Reynolds shear stress $\overline{u'v'}$ in each quadrant as
\begin{equation}
    \overline{u'v'}_j = \int_{Q_j} u'v' \, P(u', v') \, du'\, dv',
\end{equation}
where $Q_j$ denotes the $j$-th quadrant and $P(u', v')$ is the joint probability density function for an event in the $(u', v')$-space.
Moreover, one can define
\begin{equation}
    \overline{u'v'}_{1+3} = \overline{u'v'}_1 + \overline{u'v'}_3
\end{equation}
and likewise for $\overline{u'v'}_{2+4}$.

\begin{figure}[ht!]
    \centering
    \includegraphics{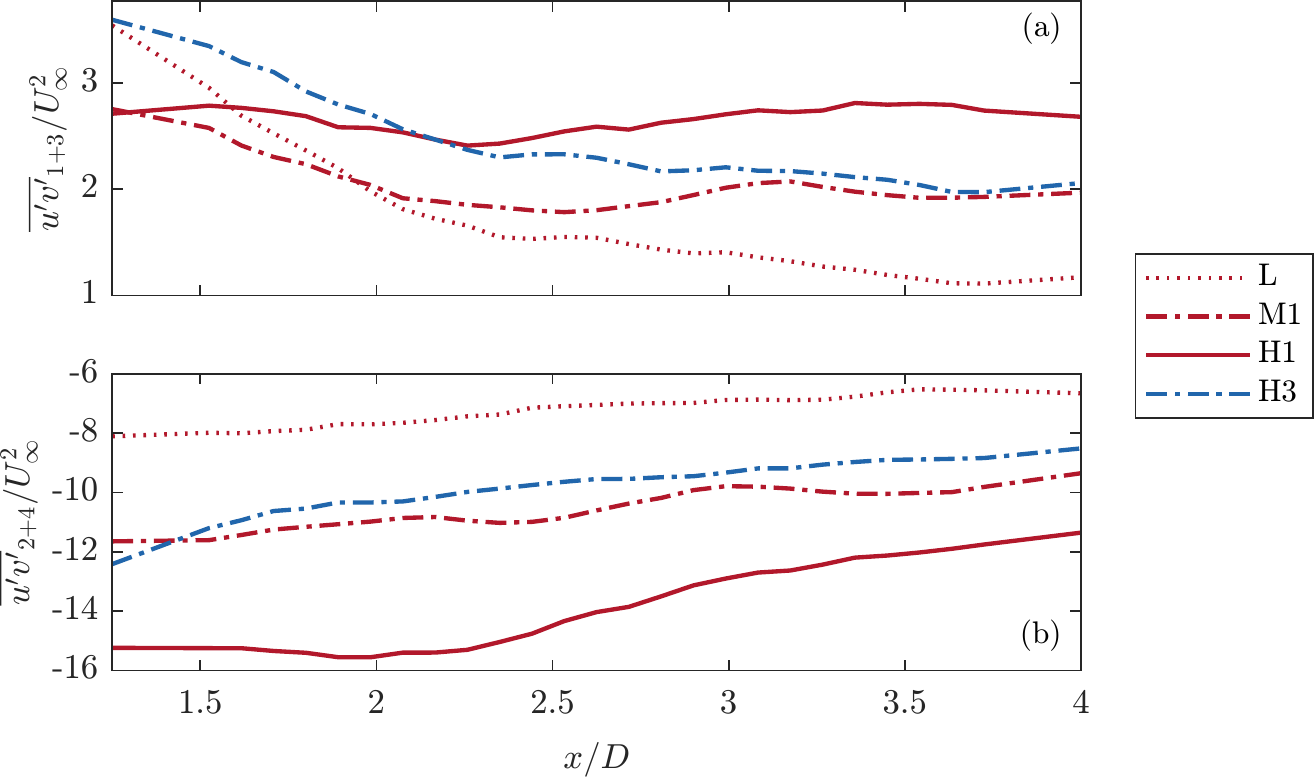}
    \caption{Sum of the contributions to the Reynolds shear stress in the first and third quadrant (\textit{a}) and in the second and fourth quadrant (\textit{b}), on the mean tip-vortices trajectory. Data for inflows L, M1, H1, and H3, and $\tsr = 3.8$.}
    \label{fig:rss-quadrant-vortices}
\end{figure}
The trends of the quadrant contributions on the mean vortices trajectory are reported in \cref{fig:rss-quadrant-vortices}, divided in events that hamper the wake evolution (\cref{fig:rss-quadrant-vortices}(a)) and those that favour it (\cref{fig:rss-quadrant-vortices}(b)).
By observing the trends of the second and fourth quadrant, for which a negative value indicates more intense sweep and ejection events and therefore a faster-evolving wake, one can see that the fast-evolving wake under flow H1 is indeed the one that results in the largest value of $\overline{u'v'}_{2+4}$ and thus in the most intense interaction between freestream and wake.
For the Kolmogorov-like flows, the intensity of these events is roughly proportional to the freestream turbulence intensity \iinf{}; this relationship does however not hold for the non-Kolmogorov test case of inflow H3, which is seen instead to have a value of Q2 and Q4 events slightly lower than those of flow M1, despite the larger \iinf{}.
By instead observing the trends of Q1 and Q3 events, it can be seen that the relationship between events intensity and \iinf{} need not necessarily hold close to the wind turbine: in fact, for small values of $x/D$, flow $L$ results in interaction events that are more intense than those developed for inflow M1 and H, despite the much less intense inflow turbulence; this however decays rapidly and by $x/D = 2$ flow L is once again the one for which events intensity is the smallest.
It is instead more interesting to observe the intensity of interaction events for flow H3, which are seen to be more intense than those of flow M1 along the whole of the near wake, and larger than those of flow H1 close to the turbine until $x/D \approx 2.25$.

From the data here presented, it is evident that the main cause of the delayed onset of wake evolution has to be found in the robustness of the train of tip-vortices, which is in turn driven by the shear layer instability.
A stable shear layer inhibits sweep and ejection events, and favours interaction events.
The effect of freestream turbulence intensity \iinf{} on the shear layer stability is non-trivial, as it has been seen that a non-Kolmogorov-like inflow with energy content skewed towards low-frequency contributions results in a more stable shear layer than a Kolmogorov-like flow.

\subsection{Far wake evolution} \label{subsec:results-farwake}
While the stability of the enveloping helical vortex structure is consistent with the near-to-far wake transition, this phenomenon does not provide an explanation for the rate at which the wake recovers, and thus for the different values of the wake recovery rate \kfit{} as seen in the previous \cref{fig:wake-fitting-parameters}.
To understand the drivers of wake evolution in the far wake region, it can be useful to approach the Reynolds-averaged Navier-Stokes equations, in particular the one that describes the evolution of the streamwise momentum component: for a steady three-dimensional flow this can be expressed as
\begin{equation}
    U \frac{\partial{}U}{\partial{}x} + 
    V \frac{\partial{}U}{\partial{}y} + 
    W \frac{\partial{}U}{\partial{}z} + 
    \frac{\partial{}\overline{u'^2}}{\partial{x}} + 
    \frac{\partial{}\overline{u'v'}}{\partial{y}} + 
    \frac{\partial{}\overline{u'w'}}{\partial{z}} = 
    -\frac{1}{\rho{}} \frac{\partial{}P}{\partial{}x},
    \label{eq:streamwise_momentum_original}
\end{equation}
where the contributions due to viscosity have been neglected as the Reynolds number is high. 
The momentum equation in the $y$-direction can be simplified \citep{Townsend1999} to the expression
\begin{equation}
    \frac{\partial{}\overline{v'^2}}{\partial{}y} + 
    \frac{\partial{}\overline{v'w'}}{\partial{}z} =
    -\frac{1}{\rho{}} \frac{\partial{}P}{\partial{}y}
    \label{eq:spanwise_momentum}
\end{equation}

\begin{figure}[ht!]
    \centering
    \includegraphics{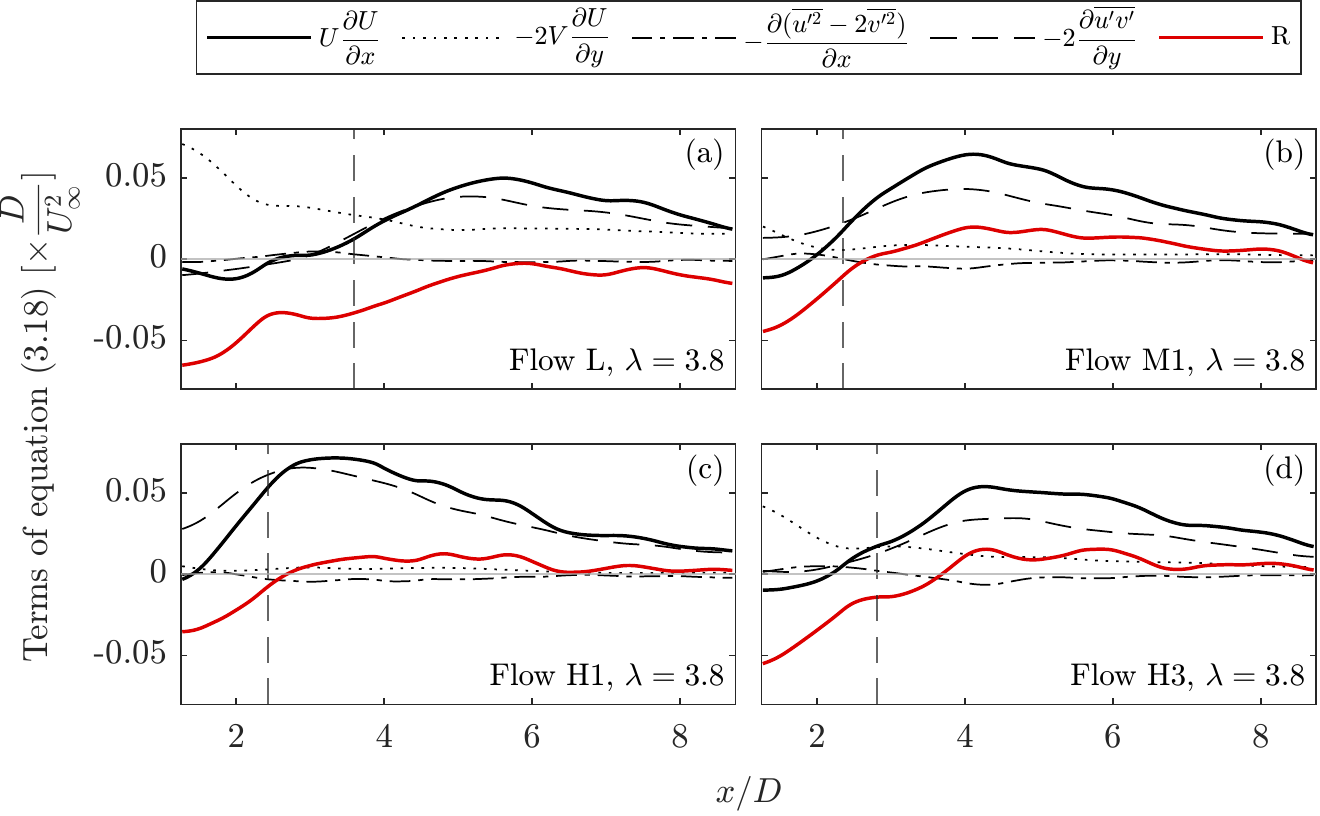}
    \caption{Terms of \cref{eq:streamwise_momentum_axsymm}, adimensionalised by multiplication with $D/\uinf^2$, for the  inflows L (\textit{a}), M1 (\textit{b}), H1 (\textit{c}), and H3 (\textit{d}), (\textit{black and red lines}), along with value of $x_0$ (\textit{dashed vertical line}); turbine operating at $\tsr = 3.8$.}
    \label{fig:streamwise-balance}
\end{figure}

As for the turbulent kinetic energy computation in \cref{subsec:results-nearwake}, one can remedy to the lack of information on the out-of-plane components of velocity by assuming that \mbox{$\frac{\partial{}}{\partial{}y} \simeq{} \frac{\partial}{\partial{}z}$} and \mbox{$\overline{v'^2} \simeq{} \overline{w'^2}$}.
While this is a strong assumption, it can be thought as reasonable especially close to the wake centreline: being the wake generated by the wind turbine roughly axisymmetric, both $y$ and $z$ represent radial directions from the wake axis outwards.
With these assumptions, \cref{eq:streamwise_momentum_original} near the wake centreline can be expressed in the simplified form
\begin{equation}
    U\frac{\partial{}U}{\partial{}x} = 
    -2 V\frac{\partial{}U}{\partial{}y}
    - \frac{\partial{}(\overline{u'^2} - 2 \overline{v'^2})}{\partial{}x}
    -2 \frac{\partial{}\overline{u'v'}}{\partial{}y} + R,
    \label{eq:streamwise_momentum_axsymm}
\end{equation}
where $R$ is a residual term that takes into account the differences between the left- and right-hand side of the equation due to the simplifying assumptions employed in this formulation, and the $-2\overline{v'^2}$ in the second term of the right-hand side derives from \cref{eq:spanwise_momentum}: with the simplifying assumption employed, one has
\begin{equation}
    2\frac{\partial{}\overline{v'^2}}{\partial{}y} = -\frac{1}{\rho{}} \frac{\partial{}P}{\partial{}y}
    \Rightarrow
    -\frac{1}{\rho{}} \frac{\partial{}P}{\partial{}x} = 2\frac{\partial{}\overline{v'^2}}{\partial{}x} +
    \frac{1}{\rho{}} \frac{\partial{}P_0}{\partial{}x},
\end{equation}
with the terms at the right-hand side of the implication sign being obtained by integrating the left-hand side in $y$ and then taking the derivative in $x$.
In particular, $P_0$ is the integration coefficient that stems from the integration of $\frac{\partial{}P}{\partial{}y}$ in $y$ and thus is constant in the vertical direction: for this reason, its evolution in $x$ is due to the presence of a potential pressure gradient due to the facility, which is arbitrarily assumed to be negligible here to simplify the analysis.
The variation of the terms of \cref{eq:streamwise_momentum_axsymm} in $x/D$ is reported, for four test cases having $\tsr = 3.8$, in \cref{fig:streamwise-balance}, along with the value of the virtual origin computed for each wake.
It can be seen that for the cases of high \iinf{} (inflows M1, H1, and H3) the dominant term of those at the right-hand side of \cref{eq:streamwise_momentum_axsymm} is indeed the one relative to the in-plane Reynolds shear stress, as it is often assumed for far wakes \citep{Tennekes1972}, and that for $x/D > x_0/D$ the residual $R$ is often small enough to be negligible; this is however not true for the case of the low \iinf{} of flow L, for which instead there is a non-negligible contribution of the mean advection term $V \frac{\partial{}U}{\partial{}y}$.
This is consistent with results previously published in literature, where higher Reynolds shear stresses generated in the turbine wake result in a swifter evolution of the streamwise velocity component \citep{Chamorro2010}, or base flows exhibiting a larger Reynolds shear stress upstream of the wind turbine result in a faster wake evolution despite similar values of free-stream \iinf{} \citep{Zhang2013}. 

Assuming therefore that, at least for the high-\iinf{} flows,
\begin{equation}
    U \frac{\partial{}U}{\partial{}x} \simeq -2 \frac{\partial{}\overline{u'v'}}{\partial{y}}
    \label{eq:streamwise_momentum_most_simplified}
\end{equation}
holds true for $x/D > x_0/D$ and on the wake centreline, one can obtain an analytical relationship between the value of the wake recovery rate $k^*$ and the in-plane Reynolds shear stress at the centreline.
In fact, noting that
\begin{equation}
    U \frac{\partial{}U}{\partial{}x} = \frac{1}{2} \frac{\partial{}U^2}{\partial{}x},
\end{equation}
one can integrate \cref{eq:streamwise_momentum_most_simplified} to yield
\begin{equation}
    \int_{x_0}^{x_\mathrm{fw}} \frac{\partial{}U^2}{\partial{}x} \, dx 
    = U_\mathrm{fw}^2 - U_{x_0}^2 
    = -4 \int_{x_0}^{x_\mathrm{fw}} \frac{\partial{}\overline{u'v'}}{\partial{y}} \, dx,
    \label{eq:integral_duvdy}
\end{equation}
where $x_\mathrm{fw}$ is a far-downstream station, $U_{x_0}$ is the value of $U$ at the virtual origin, and likewise $U_\mathrm{fw}$ is the value of $U$ downstream of the turbine.
The modified formulation of \citet{Bastankhah2014} presented in this paper can be used to estimate the values of $U$ in the wake: arbitrarily choosing $x_\mathrm{fw} = x_0 + nD$, one has
    \begin{align}
        \frac{U_\mathrm{fw}^2}{\uinf^2} &= 1 - \frac{\ct}{8(\epsilon_0 + nk^*)^2}, \\
        \frac{U_{x_0}^2}{\uinf^2} &= 1 - \frac{\ct}{8 \, \epsilon_0^2},
    \end{align}
which, substituted in \cref{eq:integral_duvdy} yield
\begin{equation}
    \frac{\ct}{8 \, \epsilon_0^2} - \frac{\ct}{8(\epsilon_0 + nk^*)^2} = -4 \int_{x_0}^{x_0+nD} \frac{1}{\uinf^2} \frac{\partial{}\overline{u'v'}}{\partial{}y} \, dx.
    \label{eq:quadratic_kstar_estimated}
\end{equation}
This is a quadratic equation in $k^*$: its positive solution is found as
\begin{equation}
    k^*_\text{est} = \frac{\epsilon_0}{n} \left( \sqrt{\frac{\ct{}}{\ct{} + 32 \, I_\text{RSS} \, \epsilon_0^2}} - 1\right)
    \label{eq:kstar_estimate}
\end{equation}
where $I_\text{RSS}$ refers to the integral of the Reynolds shear stress derivatives at the right-hand side of \cref{eq:quadratic_kstar_estimated}:
\begin{equation}
    I_\text{RSS} = \int_{x_0}^{x_0+nD} \frac{1}{\uinf^2} \frac{\partial{}\overline{u'v'}}{\partial{}y} \, dx.
\end{equation}

\begin{figure}[ht!]
    \centering
    \includegraphics{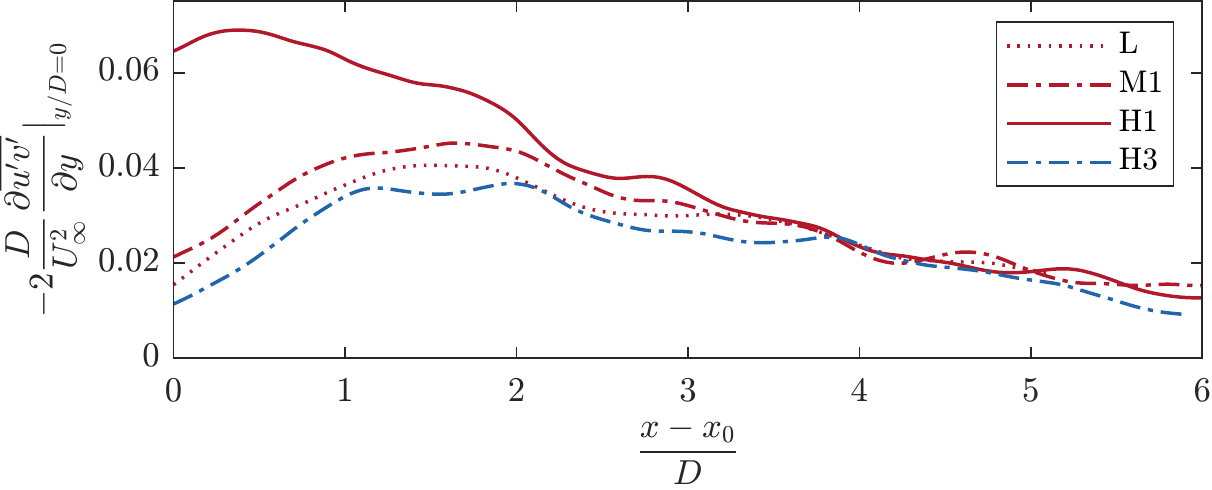}
    \caption{Distirbution of the derivative of Reynolds shear stress on the wake centreline, from the location of the virtual origin onwards. Data for $\tsr = 3.8$.}
    \label{fig:rss-centreline}
\end{figure}

\Cref{fig:rss-centreline} reports the values of the derivative of the in-plane Reynolds shear stress on the wake centreline, here approximated as the horizontal axis having $y/D = 0$.
While it can be seen that increasing the freestream turbulence intensity \iinf{} increases the Reynolds shear stress generated in the wake, the same cannot be said for the slow-evolving wake generated by inflow H3.
In fact, this last test case is seen having values of Reynolds shear stress lower than those of inflow L, despite the important mismatch in freestream turbulence between these two test cases.

\begin{figure}[ht!]
    \centering
    \includegraphics{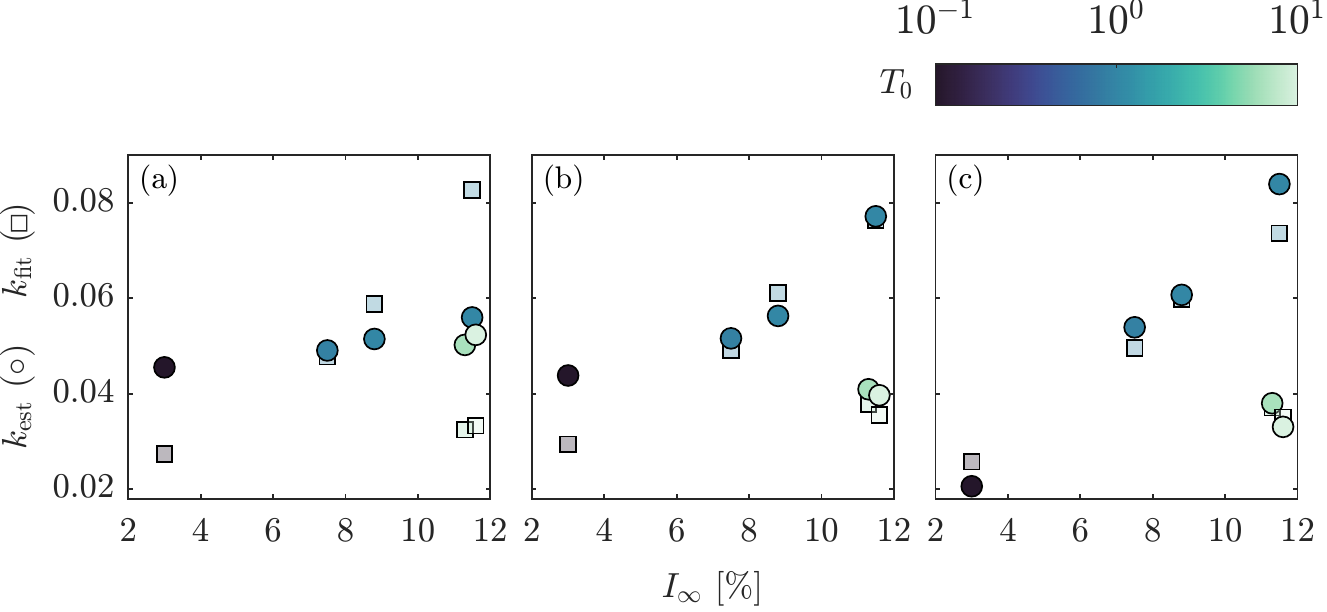}
    \caption{Estimation of \kfit{} from integration of the Reynolds shear stress on the wake trajectory according to \cref{eq:estimate-kstar-four} (\textit{round markers}) and comparison with the values found by fitting the wake velocity deficit trends (\textit{square markers}). Data from $\tsr = 1.9$ (\textit{a}), $\tsr = 3.8$ (\textit{b}), and $\tsr = 4.7$ (\textit{c}).}
    \label{fig:kstar-estimate}
\end{figure}
\Cref{fig:kstar-estimate} reports, on the top row, the trends of the estimates of $k^*$ obtained with \cref{eq:kstar_estimate}: it is evident that, with the single exception of the low-\iinf{} test case at $\tsr = 3.8$, this equation is able to capture both the linear trend of $k^*$ with \iinf{} and the drop in the wake recovery rate for the non-Kolmogorov-like inflow conditions, regardless of the thrust generated by the turbine.
However, comparison between the estimated values of $k^*$ and those reported in \cref{fig:wake-fitting-parameters} shows that \cref{eq:kstar_estimate} results in underestimations of the wake recovery rate by a factor that is approximately $4$ for the high-thrust, high-\tsr{} test cases and higher for the case of $\tsr = 1.9$
For this reason, \cref{fig:kstar-estimate}(d, e, f) reports instead on the same axis the values of \kfit{} and those of the estimated $k^*$ from integration of the Reynolds shear stress: it can be seen that the corrected equation
\begin{equation}
    \kfit \simeq 4 \frac{\epsilon_0}{n} \left( \sqrt{\frac{\ct{}}{\ct{} + 32 \, I_\text{RSS} \, \epsilon_0^2}} - 1\right)
    \label{eq:estimate-kstar-four}
\end{equation}
offers a reasonably good estimate of the wake recovery rate for the cases of high turbulence and high thrust.

Following this, it can be understood that the driver of the mean wake velocity evolution is the Reynolds shear stress: knowing this, it can be expected that flows with higher values of freestream turbulence intensity result in faster-evolving wakes as they favour a larger content of Reynolds shear stress in the wake, which explains the trend of $k^*$ with \iinf{}, seen to be linear in literature.
The non-Kolmogorov-like flows therefore cause a slower-evolving wake as they result in a lower content of Reynolds shear stress, despite similar levels of turbulent velocity fluctuations in the freestream.
A large body of literature \citep{Chamorro2015,Tobin2015,Deskos2020,Gambuzza2021} in the last years has shown a relationship between the spectrum of the incoming velocity fluctuations and the power harvested by a turbine from said incoming velocity field, with the turbine harvesting more power for spectra biased towards lower frequency.
It can therefore be assumed that, as the turbine harvests power from low-frequency velocity contributions, these will not be present in the wake: in other words, if the turbine acts as a low-pass filter from the point of view of harvested power, it must act as a high-pass filter from the point of view of its wake; this is indeed observed by \citet{Tobin2019}, in which the velocity spectra downstream of the turbine are biased towards higher-frequency contributions than those upstream of it.
A similar observation is reported by \cite{Heisel2018} for data acquired in the wake of an utility-scale wind turbine.

\begin{figure}[ht!]
    \centering
    \includegraphics{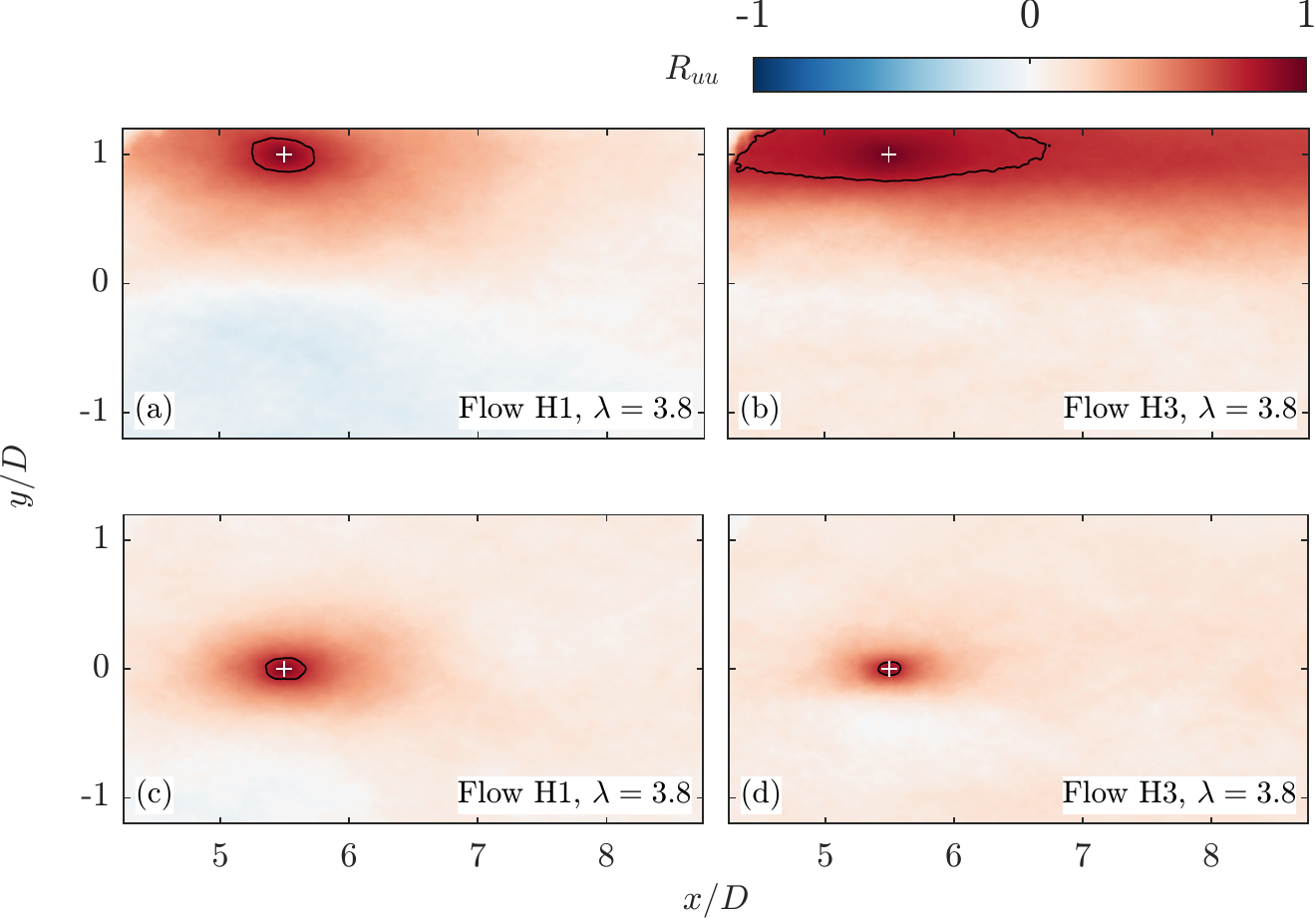}
    \caption{Two-point correlation coefficient $R_{uu}$ maps in the far wake of the turbine operating under base flows H1 (\textit{a}, \textit{c}) and H3 (\textit{b}, \textit{d}), and iso-line of $R_{uu} = 0.75$ (\textit{black line}); fixed point $x_f/D = 5, y_f/D = 1$ (\textit{a}, \textit{b}) and $x_f/D = 5, y_f/D = 0$ (\textit{c}, \textit{d}) (\textit{white plus sign}). Turbine operating at $\tsr = 3.8$.}
    \label{fig:two-pt-correlation}
\end{figure}
As no time-resolved data is available in the wake of the turbine, the hypothesis that spectra in the wake are biased towards high-frequency contributions cannot be verified; however, one can quantify the dominant scales in the wake by means of two-point correlation.
Defining the two-point correlation coefficient between a fixed point $(x_f, y_f)$ and a generic point $(x, y)$ in the domain as
\begin{equation}
    R_{uu} = \frac{\overline{u'(x_f, y_f) \, u'(x, y)}}{\left({\overline{u'^2(x_f, y_f)}} \, {\overline{u'^2(x, y)}}\right)^{1/2}},
\end{equation}
the value of $R_{uu}$ acts, in space, as analogous to the classical autocorrelation in time for a time-resolved, single-point measurement.
\Cref{fig:two-pt-correlation} reports the value of $R_{uu}$ for the far wake generated by the turbine under inflows H1, which is Kolmogorov-like, and H3, which is not; the points that are chosen as fixed are \mbox{$(x_f/D = 5.5, y_f/D = 0)$} as representative of the flow in the wake, and \mbox{$(x_f/D = 5.5, y_f/D = 1)$} for the freestream; moreover, the figure also reports the iso-line of \mbox{$R_{uu} = 0.75$} for all cases.
It can be seen that, for both the Kolmogorov-like inflow conditions of flow H1 and those of flow H3, the correlation decreases by moving from the freestream to the far wake: the characteristic lengths of the presented iso-line reduce both in $x$ and $y$ showing a bias towards lower wavelengths in the wake spectral composition, which can be thought of analogous to the bias towards high-frequency contributions observed by \citet{Tobin2019}.
Moreover, it can also be observed that, while the changes in $R_{uu}$ are limited in the vicinity of the fixed point for the case of flow H1, the difference between the correlation maps for flow H3 is instead more evident: this does further point towards a clear difference between the spectra of the freestream and the wake for non-Kolmogorov flows, as the turbine has harvested the low-frequency, high-wavelength contributions in the freestream.

To quantify this phenomenon, following the definition of \citet{Tobin2015}, one can define an ideal filter frequency
\begin{equation}
    f_\text{filt} = \frac{J \omega}{2 Q},
\end{equation}
where $J$ is the moment of inertia of all rotating parts of the turbine, $\omega$ is its angular velocity and $Q$ is the torque generated by the turbine, function of both \iinf{} and the inflow conditions.
This filter frequency defines an idealised high-pass filter $G(f)$ so that
\begin{equation}
    G(f) = \begin{cases}
    1 & f \ge f_\text{filt} \\
    0 & f < f_\text{filt}
    \end{cases}
\end{equation}
which can be applied to the power spectral distributions $\phi_u(f)$ presented previously in \cref{fig:fst-spectra} to define an equivalent, filtered freestream turbulence intensity \ifilt:
\begin{equation}
    \ifilt = \left(\frac{1}{\uinf^2}\int_0^\infty G(f) \phi_u(f) \, df\right)^{1/2}
     = \left(\frac{1}{\uinf^2}\int_{f_\text{filt}}^\infty \phi_u(f) \, df\right)^{1/2}.
     \label{eq:definition-ifilt}
\end{equation}
\begin{figure}[ht!]
    \centering
    \includegraphics{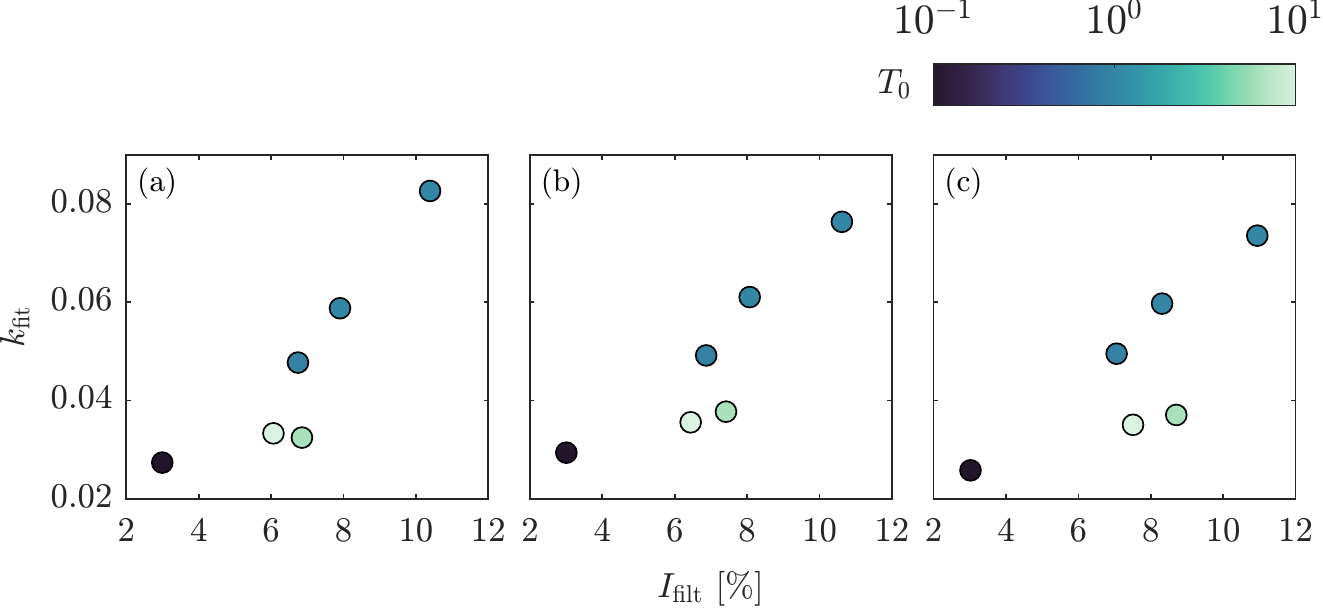}
    \caption{Trends of \kfit{} with the filtered turbulence intensity \ifilt, for $\tsr = 1.9$ (\textit{a}), $\tsr = 3.8$ (\textit{b}), and $\tsr = 4.7$ (\textit{c}).}
    \label{fig:kfit_vs_tifilt}
\end{figure}
\Cref{fig:kfit_vs_tifilt} shows the values of \kfit{} found by fitting the wake velocity deficit profiles as a function of both the freestream integral time scale and the filtered turbulence intensity \ifilt{} previously defined in \cref{eq:definition-ifilt}: note that, as \ifilt{} is a function of the turbine generated torque, this is not constant with the inflow conditions; in fact, for a given flow, this is the highest for $\tsr = 4.7$ as the ratio $\omega/Q$ decreases with $\tsr$ and therefore $f_\text{filt}$ moves towards zero with increasing $\tsr$.
It can be seen that for all cases this does not affect the linear trend of \kfit{} with freestream turbulence intensity for the Kolmogorov-like flows here analysed: this was to be expected, as the spectra of these flows are characterised by the same distribution of energy in the frequency domain.
For the case of $\tsr = 1.9$ shown in \cref{fig:kfit_vs_tifilt}(a), moreover, it can be seen that the value of \kfit{} obtained for inflow H3 at the highest \tscale{} follows the trend of \kfit{} with \ifilt{} of the Kolmogorov-like flows well, as it does the case of inflow H3 at $\tsr=3.8$; the average-\tscale{} test case of inflow H2 instead rarely follows the same trend, and neither does the case of inflow H3 for the case of $\tsr = 4.7$.
This suggests that, while the concept of the turbine acting as a high-pass filter from the point of view of the wake might be justified, the definition of the cutoff frequency for this behaviour might be different from the one that is seen in literature to regulate the power production mechanisms. Further investigation is necessary to clarify the cut-off frequency that might capture the wake effects (if any). 

\section{Conclusions} \label{sec:conclusions}
This paper has presented measurements of the wake generated by a model-scale wind turbine in a streamwise plane for streamwise distances between $1.25 D$ and $8.75 D$ downstream of the rotor-swept plane.
The freestream turbulence characteristics have been changed by means of an active grid to generate a variety of flows, of which some exhibited the canonical distribution of energy in the frequency domain of Kolmogorov, while others presented artificial contributions at very low frequencies; all flows have been defined by their freestream turbulence intensity \iinf{} and their integral time scale \tscale{}.
For Kolmogorov-like flows, it has been seen that increasing the turbulence intensity in the freestream results in a wake whose evolution starts closer to the wind turbine rotor, whose recovery is faster and whose overall length is smaller, as shown in previous literature on wakes of both wind turbines and bluff-bodies; highly-turbulent inflows moreover induce more intense wake meandering.
Non-Kolmogorov-like inflow conditions, instead, result in wakes that evolve slowly, whose evolution starts farther from the turbine and in overall longer wakes, despite the non-Kolmogorov flows here tested had the largest amount of freestream turbulence intensity here investigated; the extent of the wake meandering observed for Kolmogorov-like inflows is also reduced.

To account for these differences, analytical wake models such as the widespread Gaussian wake model \citep{Bastankhah2014} must be reformulated as a function of two parameters: a virtual origin $x_0$ and a wake recovery rate \kfit{}.
While it is customary in engineering applications to assume that \kfit{} can be inferred from the wake diameter growth rate, results reported here show this assumption might be misleading as little difference in the wake diameter is observed between different inflow conditions.
With these assumptions, namely the presence of a virtual origin and the decoupling between the wake recovery rate and the wake diameter growth, the Gaussian wake model is excellent at predicting the minimum velocity in the wake.
It has furthermore been observed that these two parameters refer to two different regions in the wake, as $x_0$ is broadly influenced by the near wake length, while \kfit{} instead describes the flow behaviour in the far wake.

In particular, the length of the near wake has been estimated in this work by considerations on the turbulent kinetic energy distribution in the turbine wake: it has been observed that the near wake length is small for high freestream turbulence intensity, and increases both with lower \iinf{} and for non-Kolmogorov-like flows, following a trend similar to that of the virtual origin $x_0$.
This trend is seen to be driven by the stability of the shear layer enveloping the wake: this is visualised as the instantaneous location of the tip-vortices shed by the turbine in its rotation, whose trajectories are considerably erratic for high-turbulence Kolmogorov inflows and more steady for low-\iinf{} and non-Kolmogorov-like inflows.
This stability or absence of it is the main driver of both the wake meandering observed in the far wake and in the length of the near wake.
It has in fact been observed that wakes whose shear layers are not stable are characterised, at the wake boundary, by intense sweep and ejection events, that favour the mixing between freestream and wake by expelling low-momentum flow from the wake and incorporating high-momentum flow from outside the wake.
This behaviour is inhibited by stable shear layers: these wakes have lower intensity of sweep and ejection events and higher intensity of interaction events, that are detrimental for the wake evolution.
In particular, close to the wind turbine, the intensity of interaction events is the largest for the non-Kolmogorov-like flow at high \iinf{} than any other wake here analysed.

The wake recovery rate in the far wake is instead seen to be dominated by the Reynolds shear stress distribution at the wake centreline: it is seen that, for all wakes here investigated except the limit case of low-\iinf{} and high power, the analytical relationship \mbox{$U \frac{\partial U}{\partial x} \simeq -2\frac{\partial \overline{u'v'}}{\partial y}$} holds true at the wake centreline.
This relationship between the mean wake velocity and the in-plane Reynolds shear stress component has been leveraged to express an analytical relationship between the wake recovery rate \kfit{} and the distribution of Reynolds shear stress; experimental data shows this relationship to hold true for the cases of high turbine thrust, while it is less accurate for the low-thrust test cases.
Non-Kolmogorov-like flows are seen to result in slower-evolving wakes as these generate a lower value of Reynolds shear stress at the wake centreline: this behaviour is connected to the phenomenon for which the turbine is shown to harvest more power form the non-Kolmogorov-like flows, thus harvesting a larger fraction of the incoming turbulent kinetic energy from the freestream.
This in literature is seen as a bias of the turbulence spectra in the wake towards higher frequency contributions; in this work, this is shown as a large difference in the two-point correlation coefficient $R_{uu}$ between freestream and wake for the non-Kolmogorov-like flow, a difference which is not evident for the classical Kolmogorov-like inflow conditions.

\section*{Funding}
This research did not receive any specific grant from funding agencies in the public, commercial, or not-for-profit sectors. The Ph.D. scholarship for author S.G. was provided by the University of Southampton.

\section*{Declaration of interests}
The authors report no conflict of interest.

\section*{Data availability statement}
The instantaneous velocity snapshots acquired during this study can be found at \url{https://doi.org/10.5258/SOTON/D2197}.

\section*{Author ORCID}
Stefano Gambuzza \url{https://orcid.org/0000-0001-9070-6901};
Bharathram Ganapathisubramani \url{https://orcid.org/0000-0001-9817-0486}

\bibliographystyle{abbrvnat}
\bibliography{biblio}

\end{document}